\documentclass[11pt]{article}
\usepackage{graphicx,mathtools,amssymb,bbm,breqn,amsxtra,amsthm,float,authblk}
\usepackage[margin=1.0in,letterpaper]{geometry}
\usepackage[title]{appendix}
\usepackage{hyperref}

\usepackage{titlesec}
\usepackage{bm}
\usepackage[textsize=tiny]{todonotes}
\titleformat*{\section}{\normalsize\bfseries}
\titleformat*{\subsection}{\normalsize\bfseries}
\titleformat*{\subsubsection}{\normalsize\bfseries}
\titleformat*{\paragraph}{\normalsize\bfseries}
\titleformat*{\subparagraph}{\normalsize\bfseries}
\newcommand{\A}{\mathbf{A}}
\newcommand{\B}{\mathbf{B}}
\newcommand{\bC}{\mathbf{C}}
\newcommand{\D}{\mathbf{D}}
\newcommand{\G}{\mathbf{G}}
\newcommand{\J}{\mathbf{J}}
\newcommand{\I}{\mathbf{I}}

\newcommand{\bfx}{\mathbf{x}}
\newcommand{\bfw}{\mathbf{w}}

\newcommand{\R}{\mathbb{R}}
\newcommand{\W}{\mathbf{W}}
\newcommand{\Ex}{\mathbb{E}}

\newcommand{\de}{\text{d}}
\newcommand{\Var}{\text{Var}}
\newcommand{\Tr}{\text{Tr}\,}

\newcommand{\bGamma}{\bm{\Gamma}}
\newcommand{\revi}[1]{\textcolor{black}{#1}}

\newtheorem{theorem}{Theorem}[section]

\theoremstyle{definition}

\theoremstyle{remark}
\newtheorem{remark}[theorem]{Remark}

\title{\Large{\textbf{Geometry-informed irreversible perturbations for accelerated convergence of Langevin dynamics}}}

\author[1]{Benjamin J.\ Zhang\footnote{Corresponding author. Email: \texttt{bjz@mit.edu}}$^\dagger$}
\author[1]{Youssef M.\ Marzouk\footnote{BJZ and YMM acknowledge support from the Air Force Office of Scientific Research, Analysis and Synthesis of Rare Events (ANSRE) MURI.}}
\author[2]{Konstantinos Spiliopoulos\footnote{KS was partially supported by the National Science Foundation (DMS 1550918, DMS 2107856) and Simons Foundation Award 672441.}}
\affil[1]{\small{Center for Computational Science and Engineering, Massachusetts Institute of Technology}}
\affil[2]{\small{Department of Mathematics and Statistics, Boston University}}

\date{\today}

\begin{document}

\maketitle

\begin{abstract}
  We introduce a novel geometry-informed irreversible perturbation that accelerates convergence of the Langevin algorithm for Bayesian computation. It is well documented that there exist perturbations to the Langevin dynamics that preserve its invariant measure while accelerating its convergence. Irreversible perturbations and reversible perturbations (such as Riemannian manifold Langevin dynamics (RMLD)) have separately been shown to improve the performance of Langevin samplers. We consider these two perturbations simultaneously by presenting a novel form of irreversible perturbation for RMLD that is informed by the underlying geometry. Through numerical examples, we show that this new irreversible perturbation can improve estimation performance over irreversible 
  perturbations that do not take the geometry into account. Moreover we demonstrate that irreversible perturbations generally can be implemented in conjunction with the stochastic gradient version of the Langevin algorithm. Lastly, while continuous-time irreversible perturbations cannot impair the performance of a Langevin estimator, the situation can sometimes be more complicated when discretization is considered. To this end, we describe a discrete-time example in which irreversibility increases both the bias and variance of the resulting estimator.
\end{abstract}

\section{Introduction}
\label{sec:intro}
Bayesian inference often requires estimating expectations with respect to non-Gaussian distributions. To solve this problem, particularly in high dimensions, one frequently resorts to sampling methods. A commonly used class of sampling methods is based on the Langevin dynamics (LD), which uses the gradient of the log-target density to specify a stochastic differential equation (SDE) whose invariant distribution is the target (e.g., posterior) distribution of interest. Long term averages over a single trajectory of the SDE can be then used to estimate expectations of interest by appealing to the ergodicity of the stochastic process. Other LD-based approaches that reduce the mean squared error (MSE) of such estimators include the Metropolis-adjusted Langevin algorithm (MALA) \cite{roberts1996exponential,girolami2011riemann}, the stochastic gradient Langevin dynamics (SGLD) \cite{welling2011bayesian}, and their variants.

It is also known that certain perturbations to the LD can accelerate convergence of the dynamics to the stationary distribution. In \cite{rey2015irreversible} the authors show that suitable reversible and irreversible perturbations to diffusion processes can decrease the spectral gap of the generator, as well as increase the large deviations rate function and decrease the asymptotic variance of the estimators.
One widely celebrated choice of \emph{reversible} perturbation is the Riemannian manifold Langevin dynamics \cite{girolami2011riemann}, in which one defines a Riemannian metric to alter the way distances and gradients are computed. The use of \emph{irreversible} perturbations to accelerate convergence has also been well studied in a variety of contexts and general settings \cite{rey2015irreversible,hwang2005accelerating,rey2015variance,rey2016improving}; see also \cite{franke2010behavior,hwang1993accelerating,Bierkens,DiaconisHolmesNeal2010} and for linear systems, \cite{lelievre2013optimal}. The authors of \cite{Ma_recipe2015} find general conditions on the drift and diffusion coefficients of an SDE so that a specified measure is the SDE's invariant measure---without, however, exploring how different choices of these coefficients impact sampling quality. 
Existing literature shows that augmenting the drift of the LD with a vector field that is {orthogonal} to the gradient of the log-target density will leave the invariant measure unchanged while decreasing the spectral gap. A convenient choice is simply to add the vector field induced by a skew-symmetric matrix applied to the gradient of the log posterior.

At the same time, traditional sampling methods for Bayesian inference are often intractable for extremely large datasets. While Langevin dynamics-based sampling methods only require access to the unnormalized posterior density, they need many evaluations of this unnormalized density and its gradient. When the dataset is extremely large, each evaluation of the density may be computationally intractable, as it requires the evaluation of the likelihood over the entire dataset. In the past decade the \emph{stochastic gradient} Langevin dynamics (SGLD) has been introduced and analyzed \cite{welling2011bayesian,teh2016consistency} to address the problem posed by large datasets. Rather than evaluating the likelihood over the entire dataset, SGLD subsamples a portion of the data (either with or without replacement) and uses the likelihood evaluated at the sampled data to estimate the true likelihood. The resulting chain can then be used to estimate ergodic averages.

In this paper we present a \emph{state-dependent irreversible} perturbation of Riemannian manifold Langevin dynamics that is informed by the \emph{geometry} of the manifold. This departs from existing literature, as the vector field of the resulting perturbation is \emph{not} orthogonal to the original drift term. This geometry-informed irreversible perturbation accelerates convergence and, if desired, can be used in combination with the SGLD algorithm to exploit the computational savings of a stochastic gradient.

We demonstrate this approach on a variety of examples: a simple anisotropic Gaussian target, a posterior on the mean and variance parameters of a normal distribution, Bayesian logistic regression, and Bayesian independent component analysis (ICA). Generally, we observe that the geometry-informed irreversible perturbation improves the convergence rate of LD compared to a standard irreversible perturbation. The improvement tends to be more pronounced as the target distribution deviates from Gaussianity. Our numerical studies also show that introducing irreversibility can reduce the MSE of the resulting long-term average estimator, mainly by reducing variance. In many cases this reduction can be significant, e.g., 1--2 orders of magnitude.

One must, however, also take the effects of discretization into account. In the continuous-time setting, it is known theoretically that irreversible perturbations can at worst only leave the spectral gap fixed. In borderline cases, though---i.e., in cases where the continuous-time theoretical improvement is nearly zero---after accounting for discretization, stiffness can actually cause the resulting estimator to perform worse than if no irreversibility were applied at all. Indeed, we will describe in Appendix \ref{sec:appendix} an illustrative Gaussian example in which the standard Langevin algorithm performs better than the algorithm with the standard irreversible perturbation---that is, an example in which additional irreversibility leads to increased bias and variance of the long term average estimator (see Remark \ref{R:NonImprovementWithIrr} and Remark \ref{R:IrrevNotImprovement} for a theoretical explanation).
Along similar lines, the idea of applying irreversible perturbations to SGLD has recently been studied in the context of nonconvex stochastic optimization \cite{hu2020non}. The authors also note that while irreversibility increases the rate of convergence, it increases the discretization error and amplifies the variance of the gradient, compared to a non-perturbed system with the same step size; see also \cite{BrosseMoulinesDurmus2018} for a related discussion on the relation of SGLD to SGD and convergence properties. This reflects the increased stiffness of irreversible SGLD relative to standard SGLD.

The rest of the paper is organized as follows. In Section \ref{sec:background} we review reversible and irreversible perturbations of the overdamped Langevin dynamics that may improve the efficiency of sampling from equilibrium. Then, in Section~\ref{sec:giirr}, we present our new geometry-informed irreversible perturbation. In Section \ref{sec:num}  we present simulation studies that demonstrate the good performance of this geometric perturbation, relative to a variety of other standard reversible and irreversible choices. In several of these examples, we also demonstrate the use of stochastic gradients.
Section~\ref{S:Conclusions} summarizes our results and outlines directions for future work. Appendix \ref{sec:appendix} details the simple Gaussian example showing that in ``borderline'' cases---i.e., when continuous-time analysis does not predict improvements from irreversible perturbations---the stiffness created by an irreversible perturbation can, after discretization, lead to poorer performance than the unperturbed case.

\section{Improving the performance of Langevin samplers}
\label{sec:background}
We begin by recalling some relevant background on Langevin samplers, Riemannian manifold Langevin dynamics, perturbations of Langevin dynamics, and the stochastic gradient Langevin dynamics algorithm. Let $f(\theta)$ be a test function on state space $E \subset \R^d$ and let $\pi(\theta)$ be some unnormalized target density on $E$. In our experiments, $\pi(\theta)$ arises as a posterior density of the form $\pi(\theta) \propto L(\theta;X) \pi_0(\theta)$, where $L(\theta; X)$ is the likelihood model, $X$ are the data, and $\pi_0(\theta)$ is the prior density. Define $\{\theta(t)\}$ as a Langevin process that has invariant density $\pi(\theta)$:
\begin{align}
	\de \theta(t) = \beta\nabla \log \pi(\theta(t)) \de t + \sqrt{2\beta}\de W(t), \label{eq:langevin}
\end{align}
where $\beta>0$ denotes the temperature, $W(t)$ is a standard Brownian motion in $\R^d$, and the initial condition may be arbitrary. By ergodicity, we may compute expectations with respect to the posterior by the long term average of $f(\theta)$ over a single trajectory:
\begin{align}
	\Ex_\pi[f(\theta)] =  \int _E f(\theta) \pi(\theta) \de \theta = \lim_{t\to\infty} \frac{1}{T} \int_0^T f(\theta(t)) \de t.  \label{eq:ergodicavg}
\end{align}
For practical computations, we must approximate \eqref{eq:ergodicavg} by discretizing the Langevin dynamics and choosing a large but finite $T$. Applying the Euler-Maruyama method to \eqref{eq:langevin} with step size $h$ yields the following recurrence relation,
\begin{align}
 	\theta_{k+1} = \theta_k + h\beta\nabla \log \pi(\theta_{\revi{k}}) \de t + \sqrt{2\beta h} \xi_{k+1}
 \end{align}
 where $\xi_{k}$ are independent standard normal random variables. The total number of steps is equal to $K = T/h$. The resulting estimator for \eqref{eq:ergodicavg} is
\begin{align}
	\Ex_\pi[f(\theta)] \approx \frac{1}{K} \sum_{k = 0}^{K-1} f(\theta_k).
\end{align}
This estimator is the \emph{unadjusted Langevin algorithm} (ULA), which has found renewed interest in the context of high-dimensional machine learning problems \cite{durmus2019high}. Discretization and truncation, however, introduce bias into the estimator. Moreover, there are noted examples in which the continuous-time process and the discretized version do not have the same invariant distribution  no matter the choice of the fixed, but nonzero, discretization step $h$; see \cite{Ganguly} for a related discussion. Certain Markov chain Monte Carlo (MCMC) methods such as MALA circumvent these issues by using the dynamics to propose new points, but accepting or rejecting them according to some rule so that the resulting discrete-time Markov chain has the target distribution as its invariant distribution \cite{roberts1996exponential,girolami2011riemann}.

Many different SDEs can have the same invariant distribution. Therefore, there has been much study into how the standard Langevin dynamics of some target distribution can be altered to increase its rate of convergence. Some examples of this can be found in the work of \cite{hwang2005accelerating,rey2016improving} and others. The standard Langevin dynamics is a reversible Markov process, meaning that the process satisfies detailed balance. The work of \cite{rey2016improving} studies, in general terms, how reversible and irreversible perturbations to reversible processes decrease the spectral gap, increase the large deviations rate function, and decrease the asymptotic variance. Yet how to \emph{choose} such perturbations to most efficiently accelerate convergence is yet to be thoroughly studied in settings beyond linear diffusion processes \cite{lelievre2013optimal}. Also, with the exception of a few examples---see for instance \cite{DuncanPavliotisZygalakis2017,lu2018analysis}---these perturbations have mainly been studied in the continuous-time setting.

\subsection{Reversible perturbations and Riemannian manifold Langevin dynamics}
We only review relevant aspects of reversible perturbations and RMLD in this section. For a detailed review of RMLD and its related Monte Carlo methods, we refer the reader to \cite{girolami2011riemann,livingstone2014information,xifara2014langevin}.  Let $\textbf{B}(\theta)$ be a $d\times d$ symmetric positive definite matrix. A reversible perturbation on LD \eqref{eq:langevin} is an SDE with multiplicative noise:
\begin{align}
	\de \theta(t) = \beta\left[\B(\theta)\nabla \log \pi(\theta(t))  + \nabla \cdot \B(\theta) \right] \, \de t + \sqrt{2\beta\B(\theta)} \de W(t).
	\label{eq:revperturb}
\end{align}
Here, the $i$-th component of $\nabla \cdot \B(\theta)$ is $\sum_{j = 1}^d \partial_{\theta_{j}} \B_{ij}(\theta)$. This is equivalent to Langevin dynamics defined on a Riemannian manifold, where the metric is $\mathbf{G}(\theta) = \B(\theta)^{-1}$ \cite{xifara2014langevin}. A straightforward calculation shows that \eqref{eq:revperturb} with $\B(\theta)$ being any symmetric positive-definite matrix admits the same invariant distribution, $\pi$. The improved rate of convergence depends on the choice of the underlying metric. The work of \cite{girolami2011riemann} argues that choosing the expected Fisher information matrix plus the Hessian of the log-prior to be the metric improves the performance of the resulting manifold MALA method. Meanwhile, \cite{rey2016improving} shows that under certain regularity conditions, if $\B(\theta)$ is chosen such that $\B(\theta) - \I$ is positive definite, then the resulting estimator is expected to have improved performance in terms of the asymptotic variance, the spectral gap, and the large deviations rate function.

\subsection{Irreversible perturbations}
Consider the following Langevin dynamics
\begin{align}
	\de \theta(t) = \left[\beta\nabla \log\pi(\theta(t)) + \gamma(\theta(t)) \right]\de t +\sqrt{2\beta} \de W(t).
\end{align}
When $\gamma(\theta) \equiv 0$, the process is reversible and has $\pi(\theta)$ as its invariant distribution. If $\gamma\neq 0$, then the resulting process will, in general, be time-irreversible unless $\gamma(\theta)$ can be written as a multiple of $\nabla \log\pi(\theta)$; see for example \cite{pavliotis2014stochastic}. However, an irreversible perturbation can still preserve the invariant distribution of the system. By considering the Fokker-Planck equation, one can show that if $\gamma(\theta)$ is chosen such that $\nabla \cdot \left( \gamma\pi \right) = 0$, then $\pi$ will still be the invariant distribution. A frequently used choice in the literature is $\gamma(\theta) = \J \nabla \log \pi(\theta)$, where $\J$ is a constant skew-symmetric matrix, i.e., $\J = -\J^T$. The computational advantage of this choice is clear since only one additional matrix-vector multiply is needed to implement it. The optimal choice of irreversible perturbation to a linear system (i.e., that which yields fastest convergence) was completely analyzed in \cite{lelievre2013optimal}.

The advantages of using irreversible perturbations is widely noted. The main result of \cite{hwang2005accelerating} is that under certain conditions, the spectral gap, i.e., the difference between the leading two eigenvalues of the generator of the Markov semigroup, increases when $\gamma \neq 0$. In \cite{rey2015irreversible,rey2015variance,rey2016improving}, the large deviations rate function is introduced as a measure of performance in the context of sampling from the equilibrium, and upon connecting it to the asymptotic variance of the long term average estimator, it is proven that adding an appropriate perturbation $\gamma$ not only increases the large deviations rate function but also decreases the asymptotic variance of the estimator. The use of irreversible proposals in the MALA was studied in \cite{ottobre2019optimal}.

\subsection{Irreversible perturbations for RMLD}
\label{sec:giirr}

In this section, we will introduce our novel geometry-informed irreversible perturbation to Langevin dynamics. Suppose that we are given a diffusion process as in \eqref{eq:revperturb}, and we want to study how to choose an irreversible perturbation that leaves the invariant distribution fixed. Indeed, our previous choice of irreversible perturbation remains valid for this system, that is, adding $\gamma(\theta) = \J \nabla \log \pi(\theta)$ for a constant skew-symmetric matrix $\J$ to the drift term of \eqref{eq:revperturb} will preserve the invariant density. This choice yields the following SDE:
\begin{align}
	\de \theta(t) = \left[(\beta\B(\theta(t))+\J) \nabla \log \pi(\theta(t)) + \beta\nabla \cdot \B(\theta(t)) \right] \de t + \sqrt{2\beta\B(\theta)} \de W(t)
	\label{eq:RMIrr}
\end{align}
We refer to this system as Riemannian manifold Langevin with an additive irreversible perturbation (\texttt{RMIrr}). This choice, however, does not take into account the relevant features that the reversible perturbation may provide when constructing an irreversible perturbation.

The reversible perturbation leads to a positive definite matrix (a metric, in the terminology of Riemannian geometry) that is state-dependent. In contrast, the skew-symmetric matrix $\J$ is fixed in the irreversible perturbation. The skew-symmetric matrix need not be constant, however, as an irreversible perturbation $\gamma(\theta)$ only needs to satisfy $\nabla \cdot (\gamma(\theta) \pi(\theta) )= 0$. In fact, if $\gamma(\theta) = \bC(\theta) \nabla \log \pi(\theta) + \nabla \cdot \bC(\theta)$ for $\bC(\theta) = -\bC(\theta)^T$, then this irreversible perturbation will also leave the invariant density intact. Noting that $\bC_{ii}(\theta) = 0$ and that $\bC_{ij} = -\bC_{ji}$, observe that
\begin{align*}
	\nabla \cdot ( \gamma(\theta) \pi(\theta)) &= \nabla \cdot ( \bC(\theta) \nabla \pi(\theta) + (\nabla \cdot \bC(\theta) ) \pi(\theta) ) \\
	 = &\sum_{i,j = 1}^d \frac{\partial \bC_{ij}(\theta)}{\partial \theta_i} \frac{\partial \pi(\theta)}{\partial \theta_j} + \bC_{ij}(\theta) \frac{\partial^2 \pi(\theta)}{\partial \theta_i \partial \theta_j} + \frac{\partial^2 \bC_{ij}(\theta)}{\partial \theta_i\partial \theta_j} \pi(\theta) + \frac{\partial \bC_{ij}(\theta)}{\partial \theta_j} \frac{\partial \pi(\theta)}{\partial \theta_i} \\
	= &\sum_{i>j, i = 1}^d \frac{\partial \bC_{ij}(\theta)}{\partial \theta_i} \frac{\partial \pi(\theta)}{\partial \theta_j} + \bC_{ij}(\theta) \frac{\partial^2 \pi(\theta)}{\partial \theta_i \partial \theta_j} + \frac{\partial^2 \bC_{ij}(\theta)}{\partial \theta_i\partial \theta_j} \pi(\theta) + \frac{\partial \bC_{ij}(\theta)}{\partial \theta_j} \frac{\partial \pi(\theta)}{\partial \theta_j} \\
	 & + \frac{\partial \bC_{ji}(\theta)}{\partial \theta_j} \frac{\partial \pi(\theta)}{\partial \theta_i} + \bC_{ji}(\theta) \frac{\partial^2 \pi(\theta)}{\partial \theta_j\partial \theta_i} + \frac{\partial^2 \bC_{ji}(\theta)}{\partial \theta_j\partial \theta_i} \pi(\theta) + \frac{\partial \bC_{ji}(\theta)}{\partial \theta_i} \frac{\partial \pi(\theta)}{\partial \theta_i} \\
	 = &\sum_{i>j, i = 1}^d \frac{\partial \bC_{ij}(\theta)}{\partial \theta_i} \frac{\partial \pi(\theta)}{\partial \theta_j} + \bC_{ij}(\theta) \frac{\partial^2 \pi(\theta)}{\partial \theta_i \partial \theta_j} + \frac{\partial^2 \bC_{ij}(\theta)}{\partial \theta_i\partial \theta_j} \pi(\theta) + \frac{\partial \bC_{ij}(\theta)}{\partial \theta_j} \frac{\partial \pi(\theta)}{\partial \theta_i} \\
	 & - \frac{\partial \bC_{ij}(\theta)}{\partial \theta_j} \frac{\partial \pi(\theta)}{\partial \theta_i} - \bC_{ij}(\theta) \frac{\partial^2 \pi(\theta)}{\partial \theta_j\partial \theta_i} - \frac{\partial^2 \bC_{ij}(\theta)}{\partial \theta_j\partial \theta_i} \pi(\theta) - \frac{\partial \bC_{ij}(\theta)}{\partial \theta_i} \frac{\partial \pi(\theta)}{\partial \theta_j}\\
	 = &\, \,  0.
\end{align*}

We seek an irreversible perturbation that takes the reversible perturbation into account, with the possibility that $\bC(\theta)$ is not a constant matrix, and investigate if it leads to any performance improvements of the long term average estimator. Note that in the literature, the above condition $\nabla \cdot (\gamma\pi)=0$ is typically rewritten into the following sufficient conditions: $\nabla \cdot \gamma(\theta) = 0$ and $\gamma(\theta) \cdot \nabla \pi(\theta) = 0$ \cite{rey2016improving}.   One can check, however, that when $\bC$ is not constant, these conditions are not met, yet $\gamma(\theta)$ is still a valid irreversible perturbation. A simple choice of $\bC(\theta)$ that incorporates $\B(\theta)$ is
\begin{align}
	\bC(\theta) = \frac{1}{2}\J \B(\theta) + \frac{1}{2}\B(\theta) \J \, ,
	\label{eq:Girrp}
\end{align}
where $\J$ is a constant skew-symmetric matrix. The $\frac{1}{2}$ factor is introduced so that if $\B(\theta) = \I$, i.e., if there is no reversible perturbation, then this perturbation reverts to the standard irreversible perturbation (\texttt{Irr}). We arrive at the following system:
\begin{align}
	\de \theta(t) = \left[ (\beta\B(\theta(t)) + \bC(\theta(t))) \nabla \log \pi(\theta(t)) + \nabla \cdot (\beta\B(\theta(t)) + \bC(\theta(t)) \right] \de t + \sqrt{2\beta\B(\theta(t))} \de W(t).
\end{align}

We call this choice of perturbation the \emph{geometry-informed irreversible perturbation} (\texttt{GiIrr}). Indeed, while there are infinitely many valid choices for $\bC(\theta)$, we will investigate the choice in \eqref{eq:Girrp} in the numerical examples. Since we will have already explicitly constructed $\B(\theta)$ and $\J$ for the other systems, the additional computational cost of computing their product will be marginal. Furthermore, as mentioned earlier, this choice reduces to \texttt{Irr} when $\B(\theta) = \I$.

One may wonder when does \texttt{GiIrr} result in improved performance over standard irreversible perturbations such as in Equation \ref{eq:RMIrr}. Based on the numerical results and intuition, we will argue that \texttt{GiIrr} results in better performance if the underlying reversible perturbation already improves the sampling. Namely, if one knows that RMLD leads to improved sampling on a given problem, then employing \texttt{GiIrr} is expected to improve sampling even further. As we mentioned earlier, the choice of \texttt{GiIrr} that is made in this paper is not unique, and a further investigation of its theoretical properties is left for future work; see also the discussion in the Section~\ref{S:Conclusions}. The goal of this paper is to present this new class of irreversible perturbations and investigate it numerically in a number of representative computational studies.

\subsection{Stochastic gradient Langevin dynamics}
In certain Bayesian inference problems, the data are conditionally independent of each other given the parameter value. Therefore, the likelihood model can often be factorized and the posterior density can be written as follows:
\begin{align}
	\pi(\theta) \propto \pi_0(\theta) \prod_{i = 1}^N \pi_i(X_i| \theta)
\end{align}
where $\pi(X_i|\theta)$ is the likelihood function for data point $X_i$.
When the dataset is extremely large, i.e., when $N\gg 1$, however, ULA becomes exceedingly expensive as it requires repeatedly evaluating the likelihood over the entire dataset for each step of the trajectory. To mitigate this challenge, the stochastic gradient Langevin dynamics was presented to reduce the computational cost of evaluating the posterior density by only evaluating the likelihood over \emph{subsets} of the data at each step. The true likelihood is estimated based on the likelihood function evaluated at the subsampled data \cite{welling2011bayesian}. Specifically, the gradient is estimated using a stochastic gradient
\begin{align}
	\nabla \log \pi(\theta|X)\approx\widehat{\nabla \log \pi(\theta|X)} = \log \pi_0(\theta) + \frac{N}{n} \sum_{i = 1}^n \log \pi(X_{\tau_i}|\theta)
\end{align}
where $\tau$ is a random subset of $\{1,\ldots, N\}$ of size $n$ drawn with or without replacement. Depending on the choice of $n$, this approach cuts down on the computational costs dramatically with some additional variance incurred by the random subsampling of the data. The original version of this algorithm made the step size variable, approaching zero as the number of steps taken $K$ became large. SGLD applied with a variable and shrinking step size was proven to be consistent: that is, the invariant distribution of the discretized system is equivalent to that of the continuous system \cite{teh2016consistency}. Having a decreasing step size counteracts the cost savings provided by computing the stochastic gradient, and therefore a version where the step size is fixed was presented in \cite{vollmer2016exploration}, where theoretical characterizations of the asymptotic and finite-time bias and variance are also developed.
In most of our numerical results, we use stochastic gradient version of the Langevin algorithm with fixed step size to demonstrate that SGLD can be used together with irreversible perturbations.

\begin{table}
\centering
\begin{tabular}{l|l|l}
 & $b(\theta)$ & $\sigma(\theta)$ \\ \hline
 \textbf{\texttt{LD}}& $\beta\nabla \log \pi(\theta)  $ & $\sqrt{2\beta} \I$  \\
 \textbf{\texttt{RM}}&  $\beta\B(\theta)\nabla \log \pi(\theta)) + \beta\nabla \cdot \B(\theta)$ & $\sqrt{2\beta\B(\theta)}$\\
 \textbf{\texttt{Irr}} & $(\beta\I + \J) \nabla \log \pi(\theta) $ & $\sqrt{2\beta} \I$ \\
 \textbf{\texttt{RMIrr}} & $(\beta\B(\theta) + \J)\nabla \log \pi(\theta) +\beta\nabla \cdot \B(\theta)$ & $\sqrt{2\beta\B(\theta)} $ \\
 \textbf{\texttt{GiIrr}} & $ (\beta\B(\theta) + \frac{1}{2}\J\B(\theta) + \frac{1}{2}\B(\theta) \J)\nabla \log \pi(\theta) + \nabla\cdot (\beta\B(\theta) + \frac{1}{2}\J\B(\theta) + \frac{1}{2}\B(\theta) \J)  $ & $\sqrt{2\beta\B(\theta)}  $
\end{tabular}
\caption{Summary of the five SDEs that share the same invariant density $\pi(\theta)$. Stochastic gradients can be considered instead of the deterministic gradients. All systems are of the form $\de \theta_t = b(\theta_t) \de t + \sigma(\theta_t) \de W_t$. The term $\beta$ denotes the temperature.}
\label{table:dynamicssummary}
\end{table}

\section{Numerical examples}
\label{sec:num}

In the following examples, we always apply the stochastic gradient version of each Langevin system unless otherwise stated. We fix $\beta = 1/2$ for all examples. The efficacy of the \texttt{GiIrr} perturbation does not change whether or not the stochastic gradient is used. We illustrate this explicitly in Section \ref{subsec:blr}, where we report the results of all perturbations both with and without the stochastic gradient, for comparison.

 \subsection{Evaluating sample quality}
 Here we discuss the measures we use to evaluate sample quality for each Langevin sampler. For each example, we estimate the \revi{bias, variance,} mean-squared error, and asymptotic variance of the estimators of the expectations of two observables: $\phi_1(\theta) = \sum_{l = 1}^d \theta^{(l)}$ and $\phi_2(\theta) = \sum_{l = 1}^d \left|\theta^{(l)}\right|^2$, where $\theta^{(l)}$ denotes the $l$th component of $\theta$. Let $\bar{\phi}^K = \frac{1}{K}\sum_{k = 0}^{K-1} \phi(\theta_k)$ denote the estimator of $\Ex_\pi[\phi(\theta)]$ obtained with a chain of length $K$.
\revi{\begin{align}
	\text{Bias}(\bar{\phi}^K) =
        \Ex\left[  \bar{\phi}^K \right] - \Ex_\pi\left[ \phi(\theta)\right]
        & =  \Ex\left[ \frac{1}{K}\sum_{k = 0}^{K-1} \phi(\theta_k)\right] - \Ex_\pi\left[ \phi(\theta)\right] \nonumber \\
      &   \approx \frac{1}{M}\sum_{i=1}^{M}\left[\frac{1}{K}\sum_{k = 0}^{K-1} \phi([\theta_k]_{i}) \right]- \Ex_\pi\left[ \phi(\theta)\right] ,
\end{align}
where $[\theta_k]_{i}$ is the state of the $i^{th}$ chain at iteration $k$.} Here, we estimate $\Ex_\pi\left[\phi(\theta) \right]$ by applying the unadjusted Langevin algorithm with a very long simulated trajectory and small discretization step.
The expected value of the estimator is computed by averaging over $M = 1000$ independent chains for the examples in Sections 3.2 and 3.3, and $M = 100$ independent chains for the examples in Sections 3.4 and 3.5. The variance of each estimator is defined and estimated as follows
\revi{\begin{align}
	\text{Var}(\bar{\phi}^K) = \Ex\left[ (\bar{\phi}^K)^2\right] - \left ( \Ex\left[\bar{\phi}^K \right] \right )^2 \approx \frac{1}{M}\sum_{i=1}^{M}\left(\frac{1}{K}\sum_{k = 0}^{K-1} \phi([\theta_k]_{i}) \right)^2 -
\left(\frac{1}{M}\sum_{i=1}^{M}\left[\frac{1}{K}\sum_{k = 0 }^{K-1} \phi([\theta_k]_{i})\right]\right)^2.
\end{align}}
\revi{The mean-squared error (MSE) of each estimator $\bar{\phi}^K$ follows analogously.}

We also evaluate the asymptotic variance of the estimator of each observable, defined as
\begin{align}
	\sigma^2(\phi) = \lim_{t\to\infty} t \, \Var\left(\frac{1}{t}\int_0^t  \phi(\theta_t) \de t\right) \approx \lim_{K\to \infty } Kh \Var\left(\bar{\phi}^K \right) .
\end{align}
To compute these asymptotic variances, we use the batch means method in \cite{asmussen2007stochastic}. After the burn-in period, we evaluate the observable over each chain. Each observable chain is then batched into twenty separate chains, and their means are evaluated. The asymptotic variance is estimated by computing the empirical variance of those means and then multiplying by the length of each of the subsampled trajectories.

In addition \revi{to measuring the performance of estimators of specific observables, for each sampler} we also evaluate overall sample quality by computing the recently-proposed kernelized Stein discrepancy (KSD) \cite{gorham2017measuring}. The KSD is a computable expression that can approximate certain integral probability metrics (IPMs) for a certain class of functions defined through the action of the Stein operator on a reproducing kernel Hilbert space. Let $\hat{\pi}_K = \frac{1}{K}\sum_{k = 0}^{K-1} \delta_{\theta_k}$ be the empirical approximation to $\pi$ based on samples $\{\theta_k\}_{k=0}^{K-1}$ produced by some Langevin algorithm. The IPM is defined as
\begin{align}
	d_{\mathcal{H}}(\hat{\pi}_K,\pi) \coloneqq \sup_{h\in \mathcal{H}} \left|\Ex_{\hat{\pi}_K}[h(Z)] - \Ex_{\pi}[h(X)] \right|,
\end{align}
for some function space $\mathcal{H}$, \revi{where $h\in \mathcal{H}$ are functions from $\R^d$ to $\R$,  and $Z\sim \hat{\pi}_K$ and $X\sim \pi$ are random variables. When $\mathcal{H}$ is large enough, $d_{\mathcal{H}}(\hat{\pi}_K,\pi)\rightarrow 0$ holds only if $\hat{\pi}_K\rightarrow\pi$ in distribution; see \cite{gorham2017measuring} and the references therein. This result implies that a better empirical approximation (i.e., better-quality samples) corresponds with a lower IPM value. Practically computing the IPM directly is intractable as it requires exactly knowing expectations with respect to the target distribution $\pi$ (which is, after all, our original goal in sampling). By judiciously choosing the function space $\mathcal{H}$, however, we can estimate the IPM by computing the KSD, using only the samples that comprise the empirical distribution $\hat{\pi}$. }

\revi{
As the expectation with respect to the target distribution $\pi$ is intractable to compute, one seeks an $\mathcal{H}$ such that for all $h\in \mathcal{H}$, $\Ex_\pi[h(X)] = 0$. Such a function space is found through \emph{Stein's identity}, which states that $\Ex_\pi[\mathcal{A}g(X)] = 0$ for all $g:\R^d\to \R^d$ in some function space $\mathcal{G}$, where $\mathcal{A}$ is the Stein operator,
\[
\mathcal{A}g(x)=\frac{1}{\pi(x)}\nabla\cdot\left( \pi(x)g(x)\right)  = g(x)\cdot\nabla\log\pi(x) + \nabla \cdot g(x),
\]
and $\nabla \cdot$ denotes the divergence operator. The space $\mathcal{G}$ is defined by specifying a reproducing kernel Hilbert space (RKHS) $\mathcal{R}_r$ of functions from $\mathbb{R}^d$ to $\mathbb{R}$, with a user-defined kernel $r(x,y)$ that is twice-continuously differentiable \cite{gorham2017measuring,izzatullah2020bayesian}. Each function $g \in \mathcal{G}$ is made up of components $g_j\in \mathcal{R}_r$, for $j = 1,\ldots,d$, such that the vector $(\|g_1 \|_{\mathcal{R}_r},\ldots,\|g_d \|_{\mathcal{R}_r})$ has unit norm in the dual space $\ell^2$, where $\|\cdot\|_{\mathcal{R}_r}$ is the norm of the RKHS \cite{gorham2017measuring}.}
\revi{Hence, if we set $h = \mathcal{A} g$ for $g \in \mathcal{G}$, then $\Ex_\pi[h(X)] = 0$. Proposition 1 in \cite{gorham2017measuring} demonstrates that such a $\mathcal{G}$ is an appropriate domain for $\mathcal{A}$ so that one indeed has $\Ex_\pi[\mathcal{A}g(X)] = 0$ for all  $g \in \mathcal{G}$.  Then the space $\mathcal{H}$ is defined as $\mathcal{H} = \mathcal{A} \mathcal{G}$, i.e., the space of functions resulting from the Stein operator applied to functions in $\mathcal{G}$. With this $\mathcal{H}$, we define the corresponding KSD of a measure $\mu$ as $\mathcal{S}(\mu)=d_{\mathcal{H}}(\mu,\pi)=\sup_{h\in \mathcal{H}}\left|\Ex_{\mu}[h(X)] \right|$.}

\revi{Proposition 2 in \cite{gorham2017measuring} shows that the KSD admits a closed form. Indeed, for any $j=1,\ldots, d$ and letting $b_{j}(x)=\partial_{x_{j}}\log\pi(x)$,
define the Stein kernel
\begin{align}
	r_0^j(x,y) &= b_j(x)b_j(y) r(x,y) + b_j(x) \nabla_{y_j}r(x,y)+b_j(y) \nabla_{x_j}r(x,y) + \nabla_{x_j}\nabla_{y_j} r(x,y).
\end{align}
Then, by Proposition 2 of \cite{gorham2017measuring}, if $\sum_{j=1}^{d}\mathbb{E}_{\mu}\left[\sqrt{r_0^j(X,X)}\right]<\infty$, we have the equality
$\mathcal{S}(\mu) = \|w\|_2$,
where for $j=1,\ldots,d$ we have
\begin{align*}
\ w_j &= \sqrt{\Ex_{\mu\times\mu}\left[r_0^j(X,\tilde{X})\right]}, \ \  X,\tilde{X} \stackrel{\text{\tiny i.i.d.}}{\sim} \mu.
\end{align*}}

\revi{Also, \cite{gorham2017measuring} emphasizes the importance of choosing the kernel $r(x,y)$ carefully so that convergence of the KSD to zero (for a sequence of empirical distributions) implies convergence in distribution to the target measure. As suggested by the results of \cite{gorham2017measuring}, we use the inverse multiquadric kernel $r(x,y) = (c^2 + \|x-y\|_2^2)^\beta$ with $\beta = -1/2$ and $c = 1$. This choice has also been used in \cite{izzatullah2020bayesian}.
}

\revi{In our examples, $\mu = \hat{\pi}_K$ is a discrete distribution, so the KSD can be computed by evaluating the kernels over all pairs of sample points \cite{gorham2017measuring}. Namely, if $\mu = \hat{\pi}_K = \frac{1}{K} \sum_{k = 0}^{K-1} \delta_{\theta_k}$, then
\begin{align}
	w_j = \sqrt{\sum_{k,k' = 1}^K r_0^j(x_k,x_{k'}) } \, .
\end{align}
We use the KSD to evaluate the quality of every Langevin sampler---effectively, sampler performance over a large class of observables---to complement our bias/variance/MSE computations for particular choices of observable. For each example below and for each Langevin sampler, we compute the KSD for 25 independent chains. The mean KSDs of the chains for each sampler are then plotted as a function of the number of steps of the chain. We comment on the quality of each sampler given these KSD plots. We refer the reader to \cite{gorham2019measuring,gorham2015measuring,gorham2017measuring,izzatullah2020bayesian}
for further details on these choices and for further literature review. }


 \subsection{Linear Gaussian example}
\label{sec:linear}

Suppose we have data $\{X_i\}_{i = 1}^N \subset \R^d$ generated from a multivariate normal distribution with mean $\theta \in \R^d$ and known precision matrix $\bGamma_X \in \R^{d\times d}$. From the data, we infer the value of $\theta$. Endow $\theta$ with a normal prior with mean zero and precision $\bGamma_\theta \in \R^{d\times d}$. Then the posterior distribution is Gaussian with mean and precision
\begin{align}
	\mu_p = (\bGamma_\theta+ N \bGamma_X)^{-1}\bGamma_X \sum_{i = 1}^N X_i\,\,\,\, \text{and} \,\,\,\, \bGamma_p = (\bGamma_\theta+N  \bGamma_X),
\end{align}
respectively. The Euler-Maruyama discretization with constant step size $h$ applied to the corresponding Langevin dynamics is
\begin{align}
	\theta_{k+1} = \left(\I - \bar{\A}h \right)\theta_k + \bar{\mathbf{D}}_k h + \sqrt{h}\xi_k
\end{align}
where
\begin{align*}
	\bar{\A} = \frac{1}{2} (\bGamma_\theta + N \bGamma_X), \,\,\,\, \bar{\mathbf{D}}_k = \frac{1}{2}\bGamma_X \sum_{i = 1}^N X_i, \,\,\,\, \xi_k \sim \mathcal{N}(0,\I).
\end{align*}
Using stochastic gradients yields the same recurrence above except with
\begin{align}
	\bar{\mathbf{D}}_k = \frac{1}{2}\bGamma_X \frac{N}{n} \sum_{i = 1}^n X_{\tau^k_i}
\end{align}
where $n \le N$ and $\tau^k_i\in\{1,\ldots, N\}$ is randomly sampled (with or without replacement) \cite{welling2011bayesian}. Expectations with respect to the posterior are approximated by an long term average of the observable over the course of a trajectory. It has been shown that despite subsampling the data at each step in the dynamics, this estimator has comparable performance as the estimator produced by the regular Langevin dynamics with the full likelihood or MALA \cite{welling2011bayesian,vollmer2016exploration}.

Now, we consider the case where the dynamics are perturbed by an irreversible term that preserves the invariant distribution of the dynamics. We demonstrate that this leads to a lower MSE than standard SGLD or Langevin dynamics. In this case, we replace $\bar{\A}$ and $\bar{\mathbf{D}}_k$ with ${\A}$ and ${\mathbf{D}}_k$, which are
\begin{align}
	{\A} = \frac{1}{2}(\I+\J)(\bGamma_\theta + N\bGamma_X), \,\,\,\, {\mathbf{D}}_k  = \frac{1}{2}(\I + \J)\bGamma_X \frac{N}{n} \sum_{i = 1}^n X_{\tau_i^k}.
\end{align}
and $\J$ is a skew-symmetric matrix.

\begin{figure}[H]
\centering
\includegraphics[width = \textwidth]{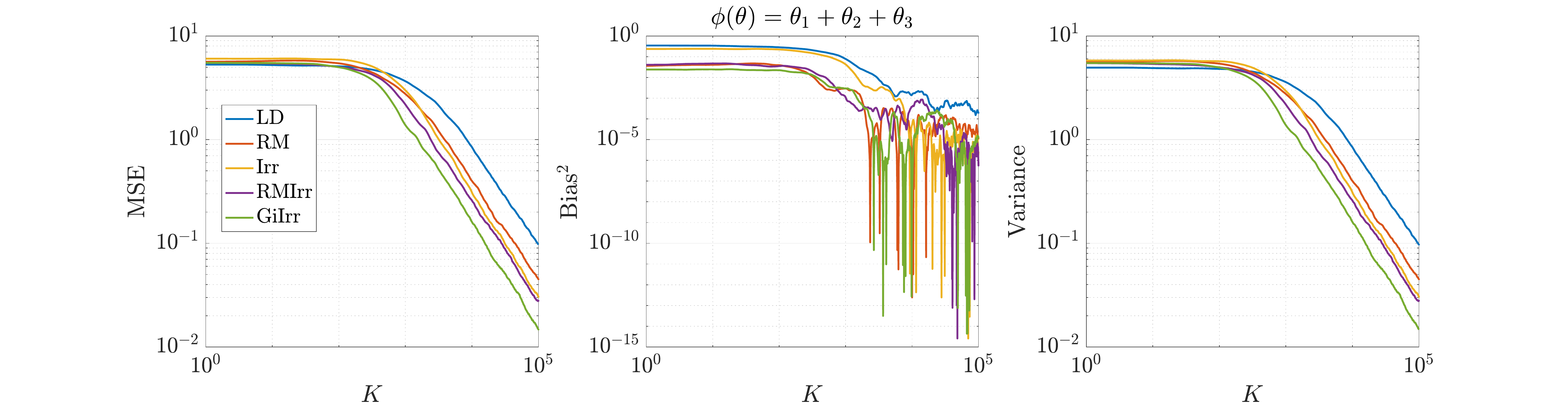}
	\caption{MSE of the running average for the first moment. Stochastic gradients are computed. }
	\label{fig:3Dgauss0}
\end{figure}

\begin{figure}[H]
\centering
\includegraphics[width = \textwidth]{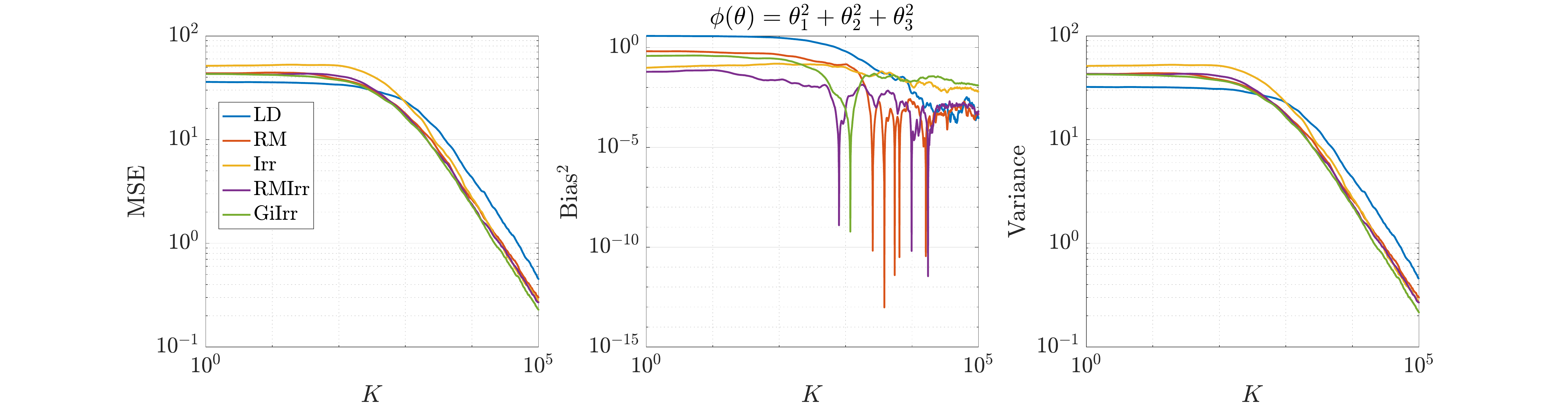}
	\caption{MSE of the running average for the second moment. Stochastic gradients are computed.}
	\label{fig:3Dgauss}
\end{figure}

For the numerical experiments, we choose $d = 3$, $N = 10$, where the mini-batches are of size $n = 2$. The initial condition is the zero vector. We have $\bGamma_X = 0.25\I$, $\bGamma_\theta$ is a precision matrix with eigenvalues $0.2,0.01,0.05$ and eigenvectors that are randomly generated, and $h = 0.005$. Note that these matrices were chosen so that the resulting reversible perturbation has eigenvalues greater than one. To construct the perturbations, we choose $\B = \bGamma_p^{-1}$ and $\J$ to be
\begin{align}
	\J = \delta \begin{bmatrix}
		0 & 1 & 1 \\ -1 & 0 & 1 \\ -1 & -1 & 0
	\end{bmatrix}
	\label{eq:signpattern}
\end{align}
for $\delta \in \R$. We consider the five different SDE systems presented in Table \ref{table:dynamicssummary} and investigate how the MSE, bias, and variance differs for each case. For this example, since a constant metric is used, the geometry-informed irreversible perturbation simply produces a different constant skew-symmetric matrix than the other irreversible perturbations. Each system is simulated for $K = 10^5$ steps with step size $h = 5\times10^{-3}$. In Figures \ref{fig:3Dgauss0} and \ref{fig:3Dgauss}, we plot the MSE of the running average for each case when the observables are the sums of the first and second moments. In Table \ref{table:lineargaussian}, we report the asymptotic variance of the estimator for each system. We see that irreversible perturbations definitely improve the performance of the estimators, although the improvement provided by the geometry-informed irreversible perturbation seems marginal over \texttt{RMIrr} when estimating the second moments. 


When the reversible perturbation is chosen such that the drift matrix is exactly the identity (for example, when the matrix is chosen to be the covariance matrix of the posterior), additional irreversibility cannot widen the spectral gap of the system. This fact can be deduced from the results of \cite{lelievre2013optimal}. The improved performance of the geometry-informed irreversible perturbation is mostly due to the fact that the norm of the corresponding skew-symmetric matrix is greater than that of simple irreversibility. Even though one can scale the skew-symmetric matrix for the other two cases to observe similar performance as geometry-informed irreversibility, \texttt{GiIrr} accomplishes that in a more systematic way.

\begin{table}[H]
\centering
\begin{tabular}{|l|l|l|l|l|}
\hline
                       & $\Ex[\text{AVar}_{\phi_1}]$    & $\text{Std}[\text{AVar}_{\phi_1}]$ & $\Ex[\text{AVar}_{\phi_2}]$    & $\text{Std}[\text{AVar}_{\phi_2}]$  \\ \hline
\texttt{LD}                     & $37.75$ & $11.94$ &$209.4$ & $84.38$  \\ \hline
\texttt{RM}                     & $20.09$ & $6.420$ & $132.8$& $49.21$       \\ \hline
\texttt{Irr}       & $15.72$ & $5.008$ &$135.4$ & $47.91$              \\ \hline
\texttt{RMIrr}  & $12.36$ & $3.937$ & $115.9$ & $ 40.10$             \\ \hline
\texttt{GiIrr}   & $\mathbf{7.444}$ & $2.336$ & $\mathbf{103.7}$ & ${36.78}$              \\ \hline
\end{tabular}
\caption{Asymptotic variance estimates for the linear Gaussian example. }
\label{table:lineargaussian}
\end{table}

\begin{figure}[H]
	\centering
	\includegraphics[width = 0.5\textwidth]{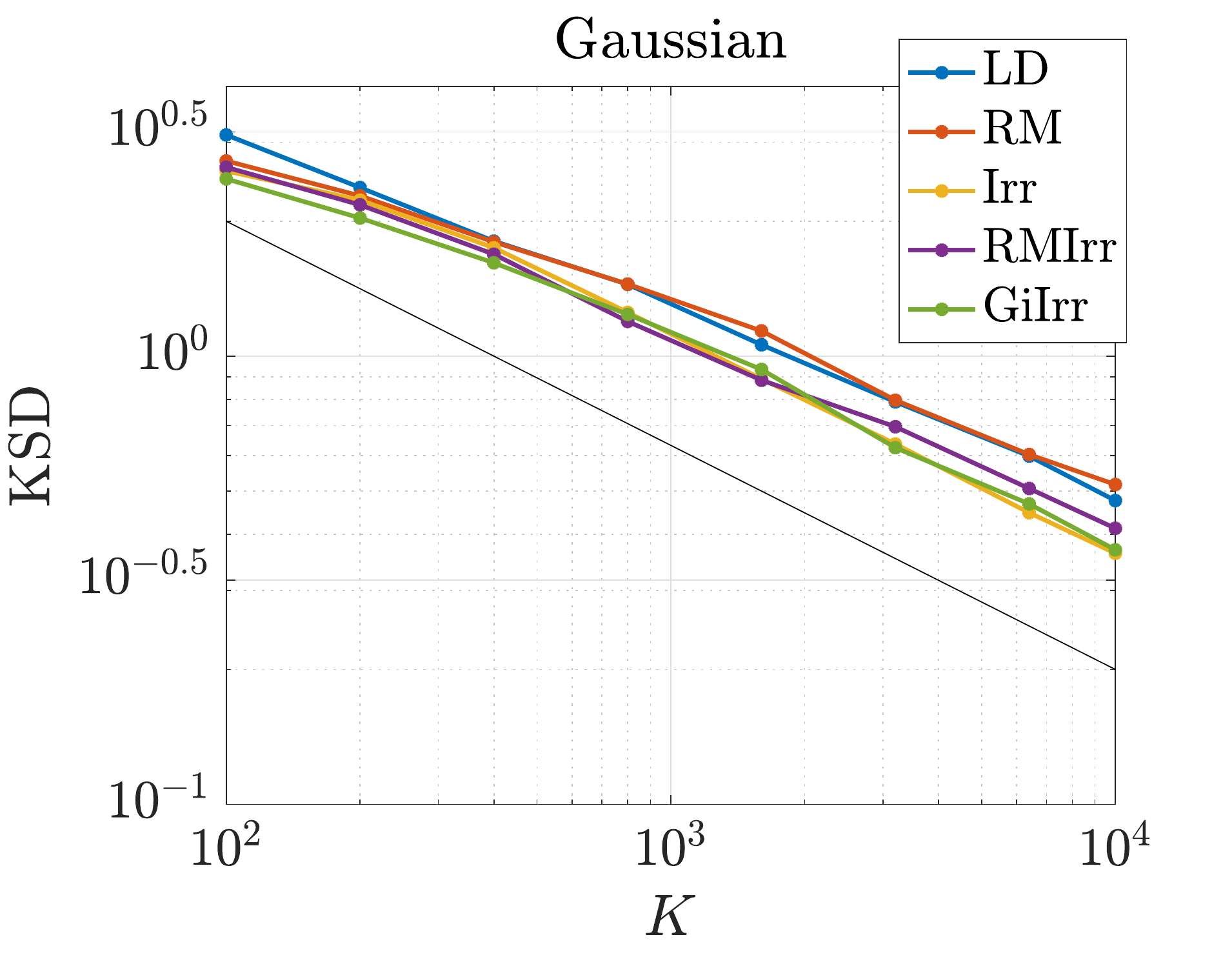}
	\caption{Kernelized Stein discrepancy plot for the Gaussian example. Black line has slope $-1/2$, which denotes the expected convergence rate.}\label{fig:KSD3Dgauss}
\end{figure}

\revi{In Figure \ref{fig:KSD3Dgauss} we plot the kernelized Stein discrepancy (KSD) for the linear--Gaussian example. We see that irreversible perturbations typically have smaller KSD compared to reversible perturbations  and that in all cases the theoretical slope of $K^{-1/2}$ (see \cite{LiuLeeJordan2016}) is achieved.}

\revi{\begin{remark}
Note that in Figure \ref{fig:KSD3Dgauss} and in all subsequent KSD-related figures, $K$ ranges up to $10^4$, rather than up to $10^{5}$ or higher (as in the bias/variance/MSE plots, e.g., Figures~\ref{fig:3Dgauss0} and \ref{fig:3Dgauss} and analogous figures in subsequent examples). The reason is that KSD is expensive to compute, and we find that evaluating it up to $K=10^4$ is sufficient to draw conclusions.
\end{remark}}

\begin{remark}\label{R:NonImprovementWithIrr}
While it is known that irreversible perturbations can, at worst, maintain the same performance as standard Langevin in the continuous-time setting \cite{rey2016improving}, when considering discretization and in borderline cases (i.e., when one does not expect much or any improvement in continuous time), irreversibility may actually harm the performance of the estimator as it introduces additional stiffness into the system without resulting in faster convergence to the invariant density. A detailed exploration of this effect is presented in Appendix \ref{sec:appendix}, in which we compute the bias and variance of the long term average estimator for a simple linear Gaussian problem where the posterior precision is a scalar multiple of the identity matrix. As further discussed in Remark \ref{R:IrrevNotImprovement}, in this case, the irreversible perturbation is not expected to lead to improvement in the sampling properties from the equilibrium. Hence, the stiffness induced upon discretization has a more profound impact on the practical performance of the irreversible perturbation.

In the current numerical study, the posterior precision is diagonal, but not a scalar multiple of the identity matrix. The eigenvalues of the resulting drift matrix are therefore distinct, and by the theory in \cite{lelievre2013optimal}, irreversible perturbations are able to reduce the spectral gap and result in improved performance. This is in contrast with the example studied in Appendix \ref{sec:appendix}. In the example of the appendix, the drift matrix is taken to be proportional to the identity matrix, and as explained in Remark \ref{R:IrrevNotImprovement}, the spectral gap therefore cannot be widened. In that case, irreversibility leads to increased stiffness of the system, which then leads to increased bias and variance in the resulting estimator.
\end{remark}

\subsection{Parameters of a normal distribution}
This example is identical to that used in \cite[Section 5]{girolami2011riemann} to demonstrate the performance of RMLD. Given a dataset of $\R$-valued data $\mathbf{X} = \{X_i\}_{i = 1}^N \sim \mathcal{N}(\mu,\sigma^2)$, we infer the parameters $\mu, \sigma$. To be clear, in this example the state is $\theta = [\mu,\sigma]^T$. The prior on $\mu,\sigma$ is chosen to be flat (and, therefore, improper). The log-posterior is
\begin{align}
	\log p(\mu,\sigma|\mathbf{X}) = \frac{N}{2} \log 2\pi - N \log \sigma - \sum_{i = 1}^N \frac{(X_i-\mu)^2}{2\sigma^2}.
\end{align}
The gradient is
\begin{align}
	\nabla \log p(\mu,\sigma|\mathbf{X}) = \begin{bmatrix}	m_1(\mu)/\sigma^2 \\ -N/\sigma + m_2(\mu)/\sigma^3
	\end{bmatrix}
\end{align}
where $m_1(\mu) = \sum_{i = 1}^N (X_i - \mu)$, and $m_2(\mu) = \sum_{i = 1}^N (X_i-\mu)^2$.
In \cite{girolami2011riemann}, the authors propose using the geometry of the manifold defined by the parameter space of the posterior distribution to accelerate the resulting Metropolis-adjusted Langevin algorithm.   The authors in \cite{girolami2011riemann} suggest using the expected Fisher information matrix to define the Riemannian metric, which in the context of reversible diffusions \cite{rey2016improving}, is equivalent to choosing $\B(\mu,\sigma)$ to be the inverse of the sum of the expected Fisher information matrix and the negative Hessian of the log-prior. Straightforward computations yield
\begin{align}
	\B = \frac{\sigma^2}{N} \begin{bmatrix}
		1 & 0 \\ 0 & 1/2
	\end{bmatrix}, \;\; \sqrt{\B} = \frac{\sigma}{\sqrt{N}} \begin{bmatrix}
		1 & 0 \\ 0 & 1/\sqrt{2}
	\end{bmatrix},\;\; \nabla \cdot \B = \begin{bmatrix}
		0 \\ \sigma/N
	\end{bmatrix}.
\end{align}
As for the geometry-informed irreversible perturbation, let $\J = \delta \begin{bmatrix}0 & 1 \\ -1 & 0 \end{bmatrix}$, for $\delta \in \R$. Then the relevant quantities are
\begin{align}
	\frac{1}{2}\J \B + \frac{1}{2}\B\J = \frac{3 \sigma^2}{4N}\J, \;\; \frac{1}{2}\nabla\cdot(\J\B + \B\J) = \frac{3\delta\sigma}{2N}\begin{bmatrix}
		1 \\0
	\end{bmatrix}.
\end{align}

\begin{figure}[H]
\centering
\includegraphics[width = \linewidth]{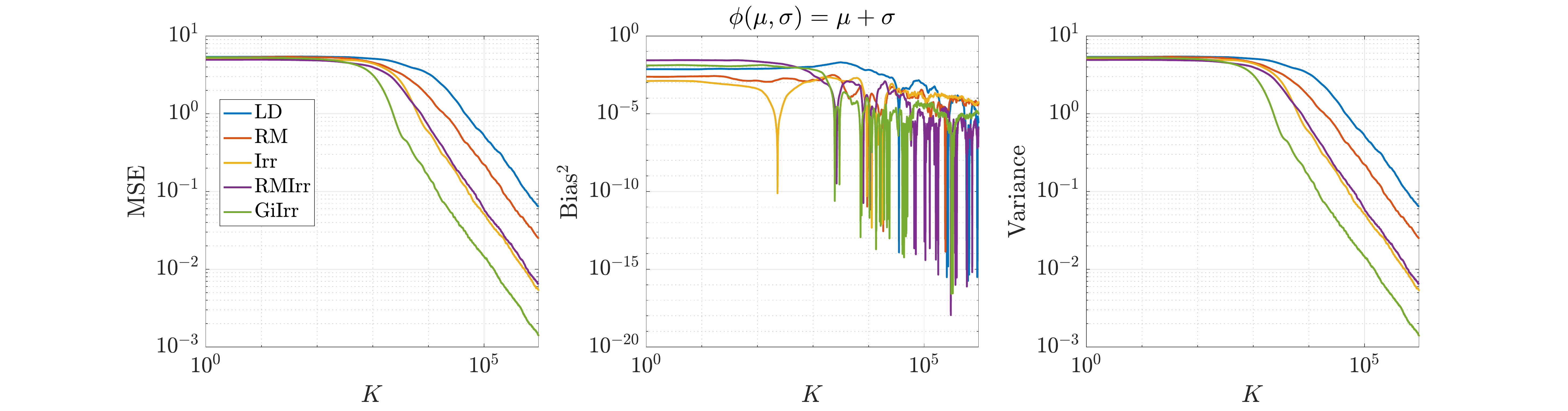}
	\caption{Observable: $\phi_1(\mu,\sigma) = \mu+\sigma$, $\delta = 2$. Stochastic gradients are computed. }
	\label{fig:1dn2sg}
\end{figure}

\begin{figure}[H]
\centering
\includegraphics[width = \linewidth]{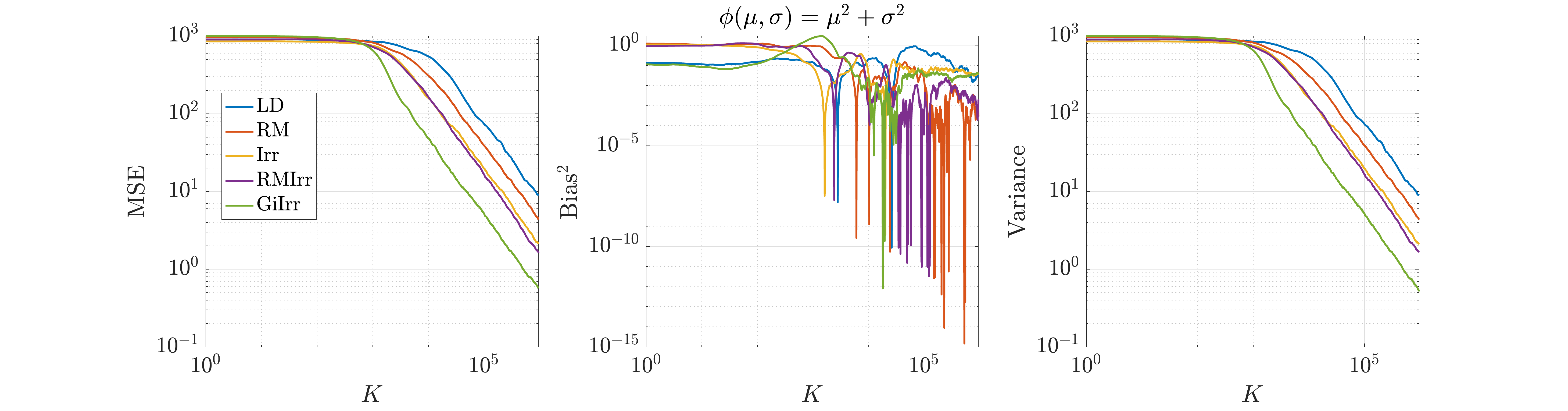}
	\caption{Observable: $\phi_2(\mu,\sigma) = \mu^2 + \sigma^2$, $\delta = 2$. Stochastic gradients are computed.}
	\label{fig:2dn2sg}
\end{figure}

\begin{figure}[H]
	\centering
\includegraphics[width = 0.5\textwidth]{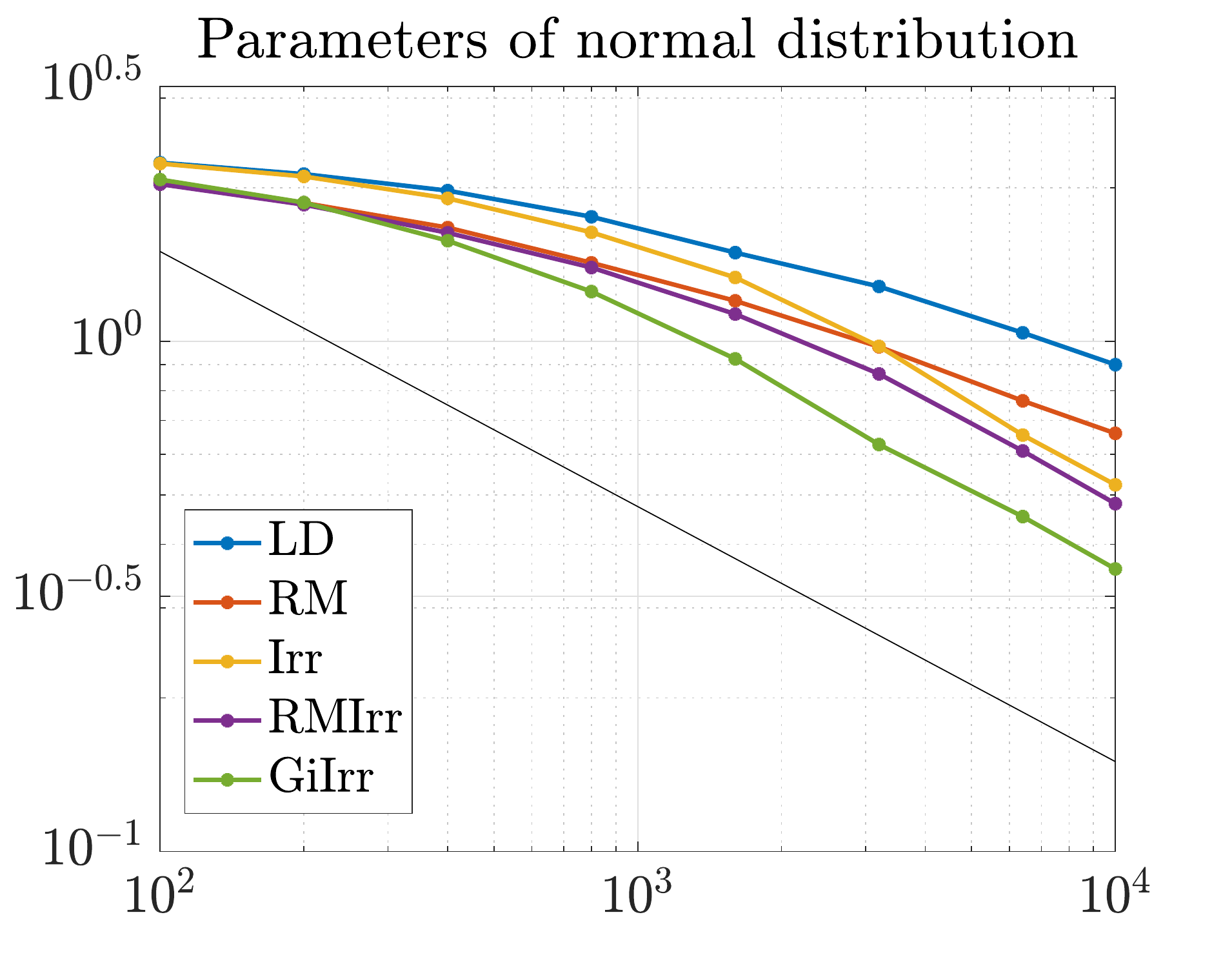}
	\caption{Kernelized Stein discrepancy plot for the parameters of a normal distribution example. Black line has slope $-1/2$, which denotes the expected convergence rate.}
	\label{fig:paramsksd}
\end{figure}

In the experiments, we have $N = 30$, $h = 10^{-3}$, $\delta = 2$. and simulate $M = 1000$ independent trajectories of each system up to $T = 1000$ for a total of $K = 10^6$ steps. The initial condition is chosen to be $\mu = 5$ and $\sigma = 20$, which is consistent with the choice in \cite{girolami2011riemann}. The data are subsampled at a rate of $n = 6$ per stochastic gradient computation. Each trajectory is allotted a burn-in time of $T_b = 10$. The dataset is generated by drawing samples from a normal distribution with $\mu_{true} = 0$ and $\sigma_{true} = 10$. The observables we study are $\phi_1(\mu,\sigma) = \mu + \sigma$, and $\phi_2(\mu,\sigma) = \mu^2 + \sigma^2$. We plot the MSE, squared bias, and variance of resulting estimators for each observable in Figures \ref{fig:1dn2sg} and \ref{fig:2dn2sg}. Moreover, in Table \ref{table:asympvarsg} we report the asymptotic variance of the estimators of each of the five systems. We plot the kernelized Stein discrepancy in Figure \ref{fig:paramsksd}. Notice that the irreversibly perturbed systems reach the $K^{-1/2}$ convergence rate (see \cite{LiuLeeJordan2016}) faster than the reversibly perturbed system. 
The main takeaway is that an irreversible perturbation that is adapted to the existing reversible perturbation performs much better than if the irreversible perturbation were applied without regard to the underlying geometry. Notice that the reversible perturbation considered here still improves the performance of the long term average estimator despite the fact that $\B -\I$ is not positive definite on the state space. Indeed, while $\B-\I$ being positive definite is a sufficient condition to obtain improved performance, it is not a necessary one \cite{rey2016improving}. The reason for the reduced asymptotic variance we observed here is because the reversible perturbation $\B$ has eigenvalues larger than one where the bulk of the posterior distribution lies.  

\begin{figure}[H]
	\centering
\includegraphics[width = 0.8\linewidth]{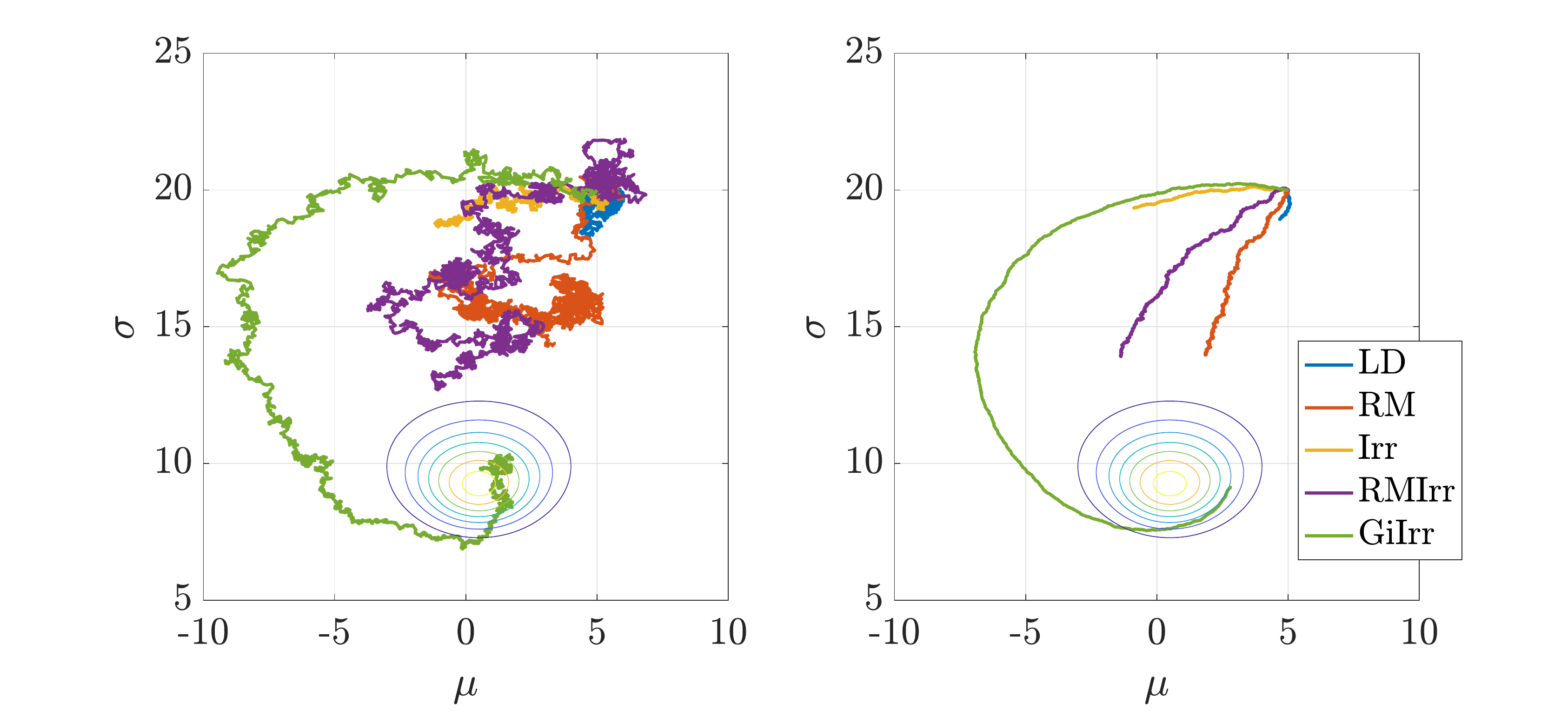}
	\caption{Trajectory burn-in: each trajectory is run for $T = 2.5$. Left: single trajectories, right: mean paths. The gradients are computed exactly here. }
	\label{fig:trajdn2}
\end{figure}

Figure \ref{fig:trajdn2} show single and mean trajectories of the burn-in period of trajectories from each of the five systems. The plot shows that the geometry-informed irreversible perturbation is able to find the bulk of the distribution sooner than the other systems without incurring additional errors due to stiffness.

To show that the \texttt{GiIrr} perturbation is not intimately tied to the stochastic gradient, we also report the results for each system when the gradients are computed exactly in Table \ref{table:asympvar}. 
We see that there is little meaningful difference in the results compared to when stochastic gradients are used.

\begin{table}[H]
\centering
\begin{tabular}{|l|l|l|l|l|}
\hline
                       & $\Ex[\text{AVar}_{\phi_1}]$    & $\text{Std}[\text{AVar}_{\phi_1}]$ & $\Ex[\text{AVar}_{\phi_2}]$    & $\text{Std}[\text{AVar}_{\phi_2}]$  \\ \hline
\texttt{\texttt{LD}}                     & $55.29$ & $21.52$ &$8332$ & $4359$  \\ \hline
\texttt{\texttt{RM}}                     & $20.63$ & $6.019$ & $4034$& $1378$       \\ \hline
\texttt{\texttt{Irr}}       & $5.791$ & $2.638$ &$2169$ & $1072$              \\ \hline
\texttt{\texttt{RMIrr}}  & $6.512$ & $2.226$ & $1729$ & $ 631.2$             \\ \hline
\texttt{\texttt{GiIrr}}   & $\mathbf{1.400}$ & $0.4697$ & $\mathbf{479.4}$ & $170.8$              \\ \hline
\end{tabular}
\caption{Asymptotic variance estimates for the parameters of a normal distribution example. Stochastic gradients are employed. }
\label{table:asympvarsg}
\end{table} 

\begin{table}[H]
\centering
\begin{tabular}{|l|l|l|l|l|}
\hline
                       & $\Ex[\text{AVar}_{\phi_1}]$    & $\text{Std}[\text{AVar}_{\phi_1}]$ & $\Ex[\text{AVar}_{\phi_2}]$    & $\text{Std}[\text{AVar}_{\phi_2}]$  \\ \hline
\texttt{\texttt{LD}} (no SG)                     & $48.51$ & $17.53$ &$7339$ & $3707$  \\ \hline
\texttt{\texttt{RM}}  (no SG)                   & $20.91$ & $6.445$ & $3855$& $1406$       \\ \hline
\texttt{\texttt{Irr}} (no SG)      & $5.658$ & $2.108$ &$2265$ & $1191$              \\ \hline
\texttt{\texttt{RMIrr}} (no SG) & $6.276$ & $2.075$ & $1648$ & $ 565.1$             \\ \hline
\texttt{\texttt{GiIrr}} (no SG)  & $\mathbf{1.363}$ & $0.4223$ & $\mathbf{492.9}$ & $183.8$              \\ \hline
\end{tabular}
\caption{Asymptotic variance estimates for the parameters of a normal distribution example. The gradients are computed exactly. }
\label{table:asympvar}
\end{table}

\revi{In Figure \ref{fig:paramsksd} we plot the KSD for the parameters-of-a-normal-distribution example. We see that \texttt{GiIrr} yields lower KSD than all other perturbations.}

\subsection{Bayesian logistic regression}
\label{subsec:blr}
Next we consider Bayesian logistic regression. Given data $\{(\bfx_i,t_i)\}_{i = 1}^N$, where $\bfx_i \in \R^d$, and $t_i \in \{0,1\}$, we seek a logistic function, parameterized by weights $\bfw \in \R^d$, that best fits the data. The weights are obtained in a Bayesian fashion, in which we endow the weights with a prior and seek to characterize its posterior distribution via sampling. Define $\varphi(y)$ to be the logistic function $\varphi(y)= ({1+\exp(-y) })^{-1}$.
The log-likelihood function is
\begin{align}
	l(\bfw) &= \sum_{i = 1}^N t_i \bfx_i^T \bfw - \sum_{i = 1}^N \log (1 + \exp(\bfx_i^T \bfw) ).
\end{align}
The prior for the weights is normally distributed with mean zero and covariance $\alpha^{-1} \mathbf{I}$. The gradient of the log-posterior is
\begin{align}
	\nabla_{\bfw} \log \pi(\bfw | \mathbf{X}) = -\alpha\bfw + \sum_{i = 1}^N t_i \bfx_i - \sum_{i = 1}^N \varphi(\bfx_i^T \bfw) \bfx_i.
\end{align}
This term is used in the drift part of the Langevin dynamics that fully computes the gradient of the log-likelihood at every step. If the data are subsampled as in SGLD, we instead compute
\begin{align}
	\nabla_{\bfw}\log \tilde{\pi}(\bfw|\mathbf{X}) = -\alpha\bfw + \frac{N}{n}\sum_{i = 1}^n t_{\tau_i} \bfx_{\tau_i} - \frac{N}{n}\sum_{i = 1} ^n \phi(\bfx_{\tau_i}^T\bfw) \bfx_{\tau_i}.
\end{align}
We use the \texttt{german} data set described in \cite{gershman2012nonparametric} for the numerical experiments. In this problem, there are 20 weight parameters to be learned. The training dataset is of size $N = 400$ and we choose to subsample at a rate of $n = 10$ per likelihood computation. The time step we choose is $h = 10^{-4}$ and $ K = 4\times10^{5}$ steps. The initial condition is chosen to be the zero vector. We generate the skew-symmetric matrix by constructing a lower triangular matrix with entries randomly drawn from $\{1,-1\}$ and then subtracting its transpose. The diagonal is then set to zero and the matrix is scaled to have norm one.

\begin{figure}[H]
\centering
\includegraphics[width = \textwidth]{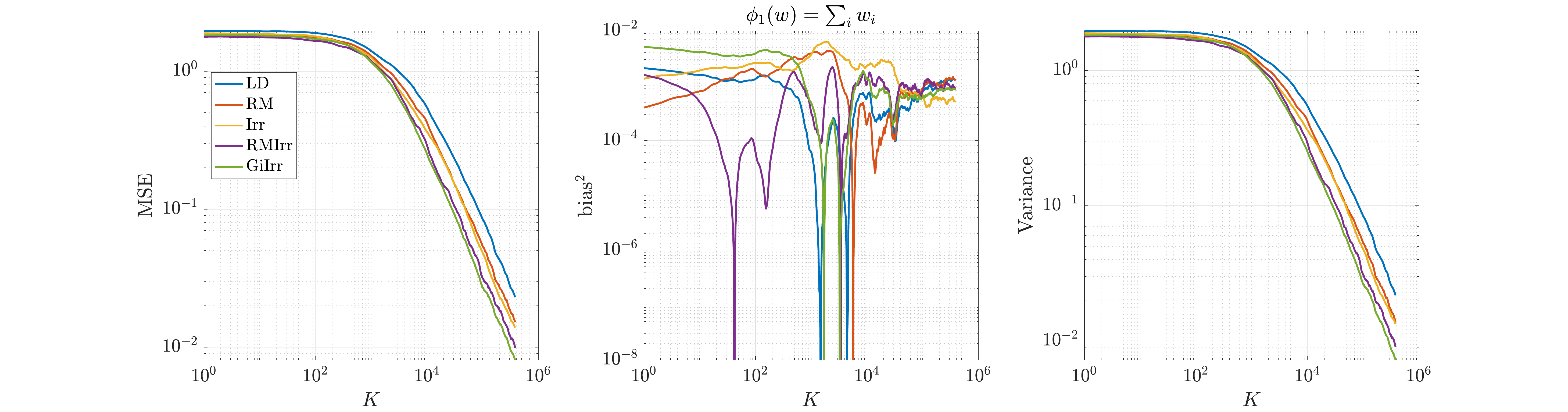}
	\caption{Observable: $\phi_1(w) = \sum_{i}w_{i}$. Bayesian logistic regression. Here, $d = 20$. }
	\label{fig:blrvariable0}
\end{figure}

\begin{figure}[H]
\centering
\includegraphics[width = \textwidth]{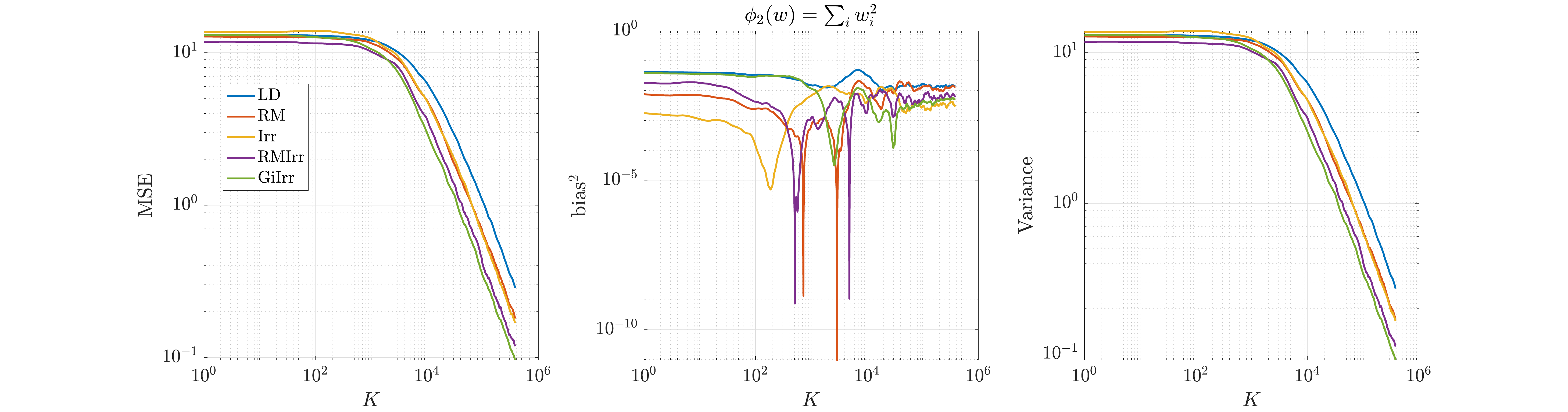}
	\caption{Observable: $\phi_2(w) = \sum_{i}w^{2}_{i}$. Bayesian logistic regression. Here, $d = 20$. }
	\label{fig:blrvariable}
\end{figure}

As for the Riemannian manifold Langevin dynamics, in \cite{girolami2011riemann} the authors use the expected Fisher information matrix plus the negative Hessian of the log-prior as the underlying metric, which in this case is equal to
\begin{align}
	\mathbf{G}(w) = \alpha^{-1}\I + \mathbf{X}\bm{\Lambda}(w) \mathbf{X}^T
\end{align}
where $\bm{\Lambda}$ is a diagonal matrix with entries $\bm{\Lambda}_{ii}(w) = (1-\varphi(\mathbf{x}_i^T w) ) \varphi(\mathbf{x}_i^Tw)$ and $\mathbf{x}_i$ is the $i$-th column of $\mathbf{X}$. The resulting reversible perturbation uses the inverse of $\G(w)$. This perturbation, however, does not lead to accelerated convergence to the invariant measure since the eigenvalues of $\mathbf{G}$ are large. This implies that the eigenvalues of $\mathbf{G}^{-1}$ are less than one and so $\mathbf{G}^{-1}(w) -\I$ is not positive definite, a condition that needs to be satisfied to guarantee accelerated convergence \cite{rey2015irreversible}. To alleviate this issue, we consider the reversible perturbation $\B(w) = \I + \G^{-1}(w)$. This guarantees $\B(w)$ to be positive definite for all $w$, but the drawback is that computing the square root of $\B(w)$ requires explicitly computing or at least approximating the inverse of $\G(w)$ repeatedly in the simulation (and not just computing the action of the inverse). This additional computational cost is incurred for all examples that consider a geometry-informed perturbation, both reversible and irreversible.
We show the result of this state-dependent perturbation in Figures \ref{fig:blrvariable0} and \ref{fig:blrvariable} and report the asymptotic variance in Table \ref{table:blr}. The geometry-informed irreversible perturbation does provide improvement over all other perturbations. We observe that the asymptotic variance is reduced by half over \texttt{RM}, with only little additional computational effort. Most of the computational cost of applying \texttt{GiIrr} is due to the evaluation of the reversible perturbation. Therefore we emphasize that if one is already applying the reversible perturbation to the Langevin dynamics, the marginal cost of applying the \texttt{GiIrr} perturbation is negligible.


\begin{table}[H]
\centering
\begin{tabular}{|l|l|l|l|l|}
\hline
                      & $\Ex[\text{AVar}_{\phi_1}]$    & $\text{Std}[\text{AVar}_{\phi_1}]$ & $\Ex[\text{AVar}_{\phi_2}]$    & $\text{Std}[\text{AVar}_{\phi_2}]$  \\ \hline
\texttt{\texttt{LD}}                     & $1.967$ & $0.9995$ &$23.77$ & $12.52$  \\ \hline
\texttt{RM}                     & $1.328$ & $0.6538$ & $15.35$& $7.348$       \\ \hline
\texttt{Irr}       & $1.163$ & $0.5698$ &$14.84$ & $7.738$              \\ \hline
\texttt{RMIrr}  & $0.8775$ & $0.4228$ & $10.68$ & $ 5.306$             \\ \hline
\texttt{GiIrr}   & $\mathbf{0.7148}$ & $0.3450$ & $\mathbf{8.798}$ & ${4.490}$              \\ \hline
\end{tabular}
\caption{Asymptotic variance estimates for the Bayesian logistic regression example with a state-dependent metric. }
\label{table:blr}
\end{table}

\begin{figure}[H]
	\centering
\includegraphics[width = 0.5\textwidth]{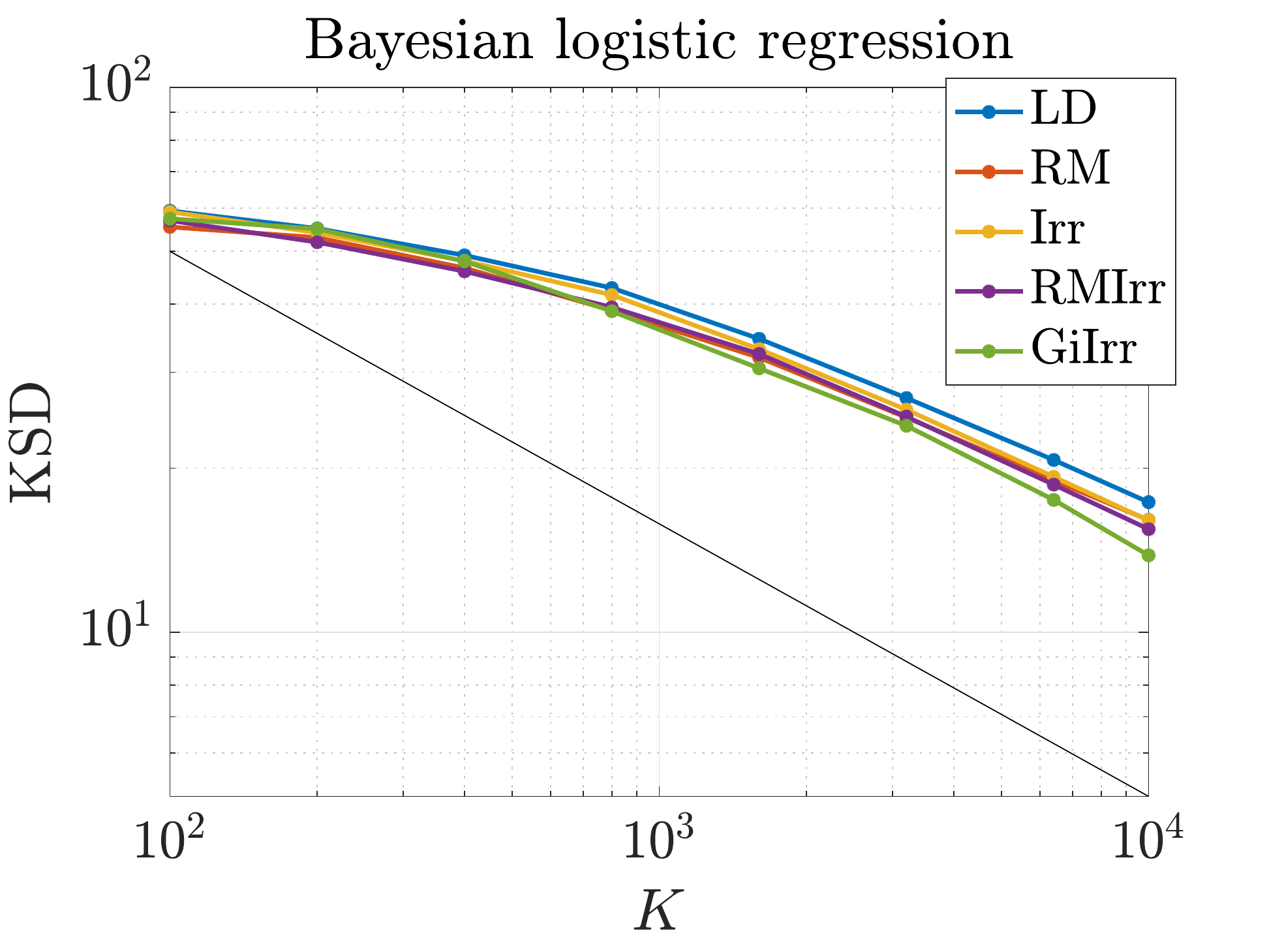}
	\caption{Kernelized Stein discrepancy plot for Bayesian logistic regression example. Black line has slope $-1/2$, which denotes the expected convergence rate.}\label{fig:KSDBayesianlogistic}
\end{figure}

\revi{In Figure \ref{fig:KSDBayesianlogistic} we plot the KSD for the Bayesian logistic regression example. We see that \texttt{GiIrr} has slightly lower KSD than the other perturbations, but the differences are small. The theoretical slope of $K^{-1/2}$ is still realized.}

\subsection{Independent component analysis}

\label{subsec:ica}
Our last example considers the problem of blind signal separation addressed in \cite{welling2011bayesian} and \cite{amari1996new}. This problem yields a posterior that is strongly non-Gaussian and multi-modal, and we show that \texttt{GiIrr} has substantially better sampling performance over standard reversible and irreversible perturbations. Suppose there are $m$ 
 separate unknown independent signals $s^i(t)$ for $i = 1,\ldots, m$ that are mixed by mixing matrix $\mathbf{M}\in \R^{d\times d}$. Suppose we can observe the mixed signals $X(t) = \mathbf{M} s(t)$ for $N$ instances in time. The goal of independent component analysis is to infer a de-mixing matrix $\mathbf{W}$ such that the $m$ signals are recovered up to a nonzero constant and permutation. As such, this problem is generally ill-posed, but is suitable to be considered in a Bayesian context. The ICA literature states that, based on real-world data, it is best to assume a likelihood model with large kurtosis. Following \cite{welling2011bayesian,amari1996new}, let
$ 	p(y_i) = \frac{1}{4} \text{sech}^2\left( \frac{1}{2} y_i\right). $
 The prior on the weights $\mathbf{W}_{ij}$ is Gaussian with zero mean and precision $\lambda$. The posterior is equal to
 \begin{align}
 	p(\W|X) \propto |\det \W| \prod_{i = 1}^m p(\mathbf{w}_i^T \mathbf{x}) \prod_{ij} \mathcal{N}(\W_{ij}; 0,\lambda^{-1}).
 \end{align}
 The gradient of the log posterior with respect to the matrix $\W$ is then
 \begin{align}
 	f(\W) = \nabla_\W \log p(\W|X) = \left( N(\mathbf{W}^T)^{-1} - \sum_{n = 1}^N \tanh\left( \frac{1}{2} \mathbf{y}_n \right) \mathbf{x}_n^T \right) - \lambda \W.
 \end{align}
 It is suggested in \cite{amari1996new} that the natural gradient should be used instead of the gradient we see here above to account for the information geometry of the problem. Specifically, \cite{teh2016consistency,amari1996new} post-multiply the gradient by $\W^T\W$ and arrive at the so-called natural gradient of the system
 \begin{align}
 	\mathcal{D}_{\mathbf{W}} \coloneqq  \left( N\I_d - \sum_{n = 1}^N \tanh\left( \frac{1}{2} \mathbf{y}_n \right) \mathbf{y}_n^T \right)\W - \lambda \W\W^T\W.
 \end{align}
 In the context of RMLD, this is equivalent to perturbing the system with a reversible perturbation with $\tilde{\B}(\W) = \mathbf{W}^T\mathbf{W}\otimes \I_d$ pre-multipled in front of the vectorized gradient. That is, we have
 \begin{align*}
 	\text{vec}(f(\mathbf{W})\mathbf{W}^T\mathbf{W}) = (\mathbf{W}^T\mathbf{W} \otimes \I_d) \text{vec}f(\mathbf{W}).
 \end{align*}
 This choice of reversible perturbation, however, may not be sufficient for accelerating the convergence of Langevin dynamics as $\tilde{\B}(\W)-\mathbf{I}_{d^2}$ is not positive definite throughout the state space \cite{rey2016improving}. Instead, we choose the reversible perturbation $\B(\W) = \I_{d^2} + (\W^\top\W \otimes \I_d) = ((\I_d + \W^\top \W)\otimes \I_d). $

We construct the \texttt{GiIrr} term as follows. To take advantage of the matrix structure of the reversible perturbation, we choose the skew-symmetric matrix such that it acts within the computation of the natural gradient. We choose $\mathbf{J} = (\I_d \otimes \bC_0) + (\bC_0 \otimes \I_d)$ where $\bC_0$ has the same sign pattern as \eqref{eq:signpattern} but such that $\mathbf{J}$ has matrix norm equal to 1. Then the geometry-informed irreversible perturbation is
\begin{align*}
 	\frac{1}{2}\B(\W)\mathbf{J} + \frac{1}{2}\mathbf{J} \B(\W) = ((\I_d+\W^T\W )\otimes \bC_0)+(\bC_0 \otimes \I_d) + \frac{1}{2}(\W^T\W \bC_0 \otimes \I_d) + \frac{1}{2}(\bC_0 \W^T\W \otimes \I_d).
 \end{align*}
To simulate the \texttt{RM} and \texttt{GiIrr} systems, correction terms (such as $\nabla \cdot \B(\theta)$) need to be computed. The correction terms are derived using the symbolic algebra toolbox in MATLAB. Since the perturbations are vectors of polynomials, the symbolic algebra toolbox can easily derive and efficiently evaluate the correction terms.

\begin{figure}
\centering
	\includegraphics[width = 0.8\textwidth]{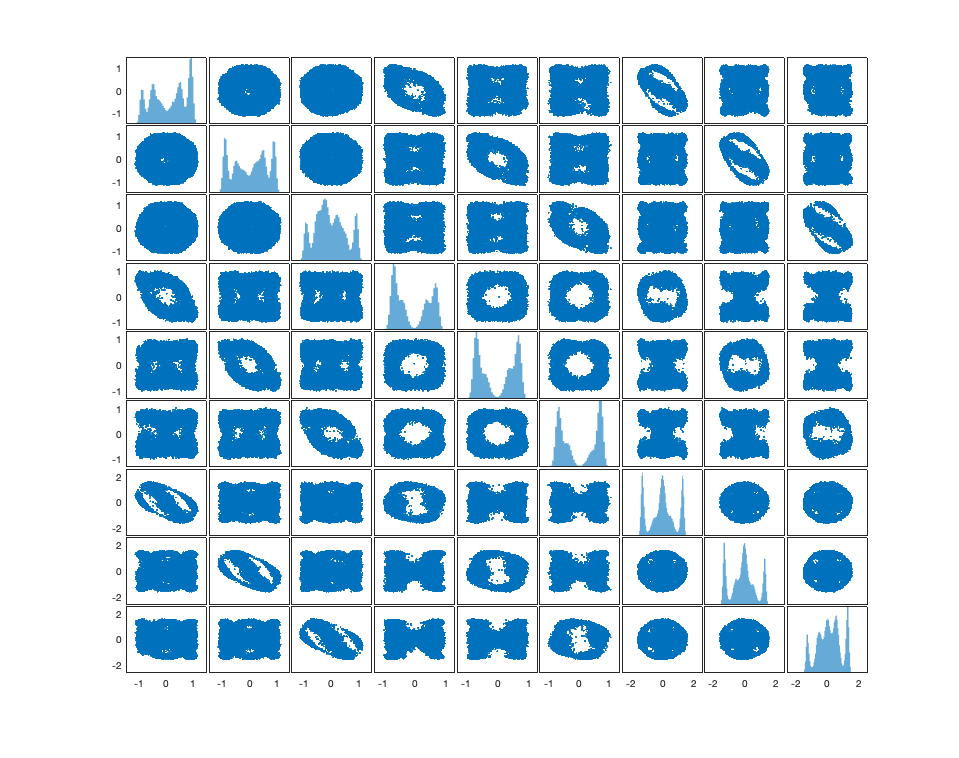}
	\caption{Posterior distribution sampled with standard Langevin with a deterministic gradient with $T = 10000$ and $h = 10^{-4}$. Notice that the system is very multimodal and non-Gaussian.}
	\label{fig:baselinepos}
\end{figure}

\begin{figure}
 	\centering
 \includegraphics[width = \textwidth]{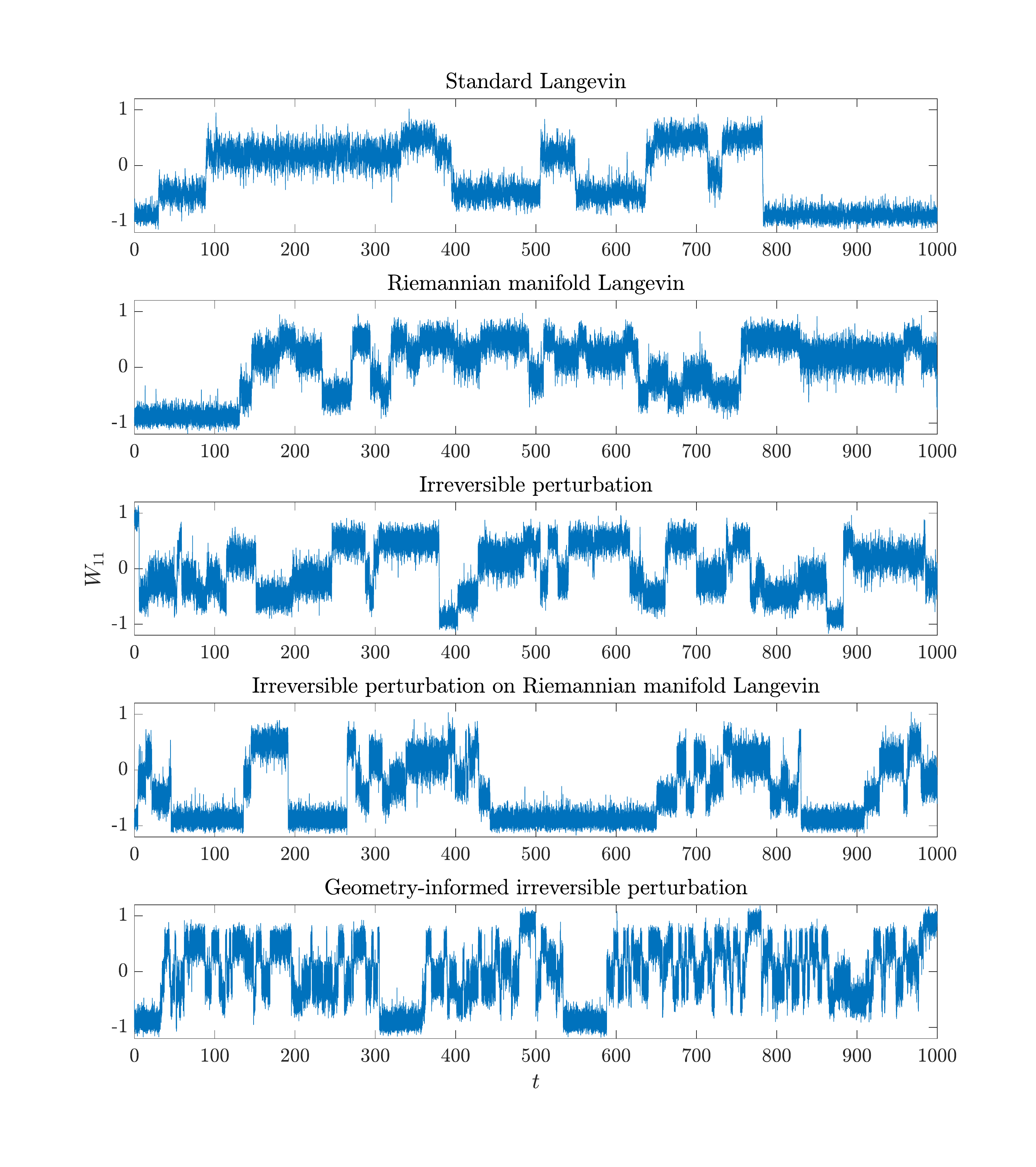}
 	\caption{Trace plots of the $W_{11}$ marginal.}
 	\label{fig:traceplots}
 \end{figure}

For the numerical experiments, we synthetically generate $m = 3$ signals, one of which is Laplace distributed, and two are distributed according to the squared hyperbolic secant distribution. The posterior distribution is $d = 9$ dimensional, there are a total of $N = 400$ data points, and the gradient is approximated by subsampling $n = 40$ data points per estimate. The initial condition here is chosen to be a diagonal matrix with either $+1$ or $-1$ entries, which are chosen randomly. Since the posterior is nine-dimensional and highly multimodal, it is difficult to evaluate its marginal densities directly, i.e., without sampling. Instead, we establish a baseline reference density by simulating the standard Langevin dynamics with exact computation of the likelihood over all the data over $T = 10000$ with $h = 10^{-4}$. One- and two-dimensional marginals of this baseline posterior distribution are plotted in Figure \ref{fig:baselinepos}. The two-dimensional marginals highlight the challenges of sampling from this posterior.
In Figure \ref{fig:traceplots}, we plot trace plots of the $\W_{11}$ variable for each system. By visual inspection, we see that that mixing is best for the geometry-informed irreversibly perturbed system. One can intuitively expect that with better mixing, the geometry-informed irreversibility should yield better estimation performance than the other systems. We assess this quantitatively below.

\begin{figure}
 	\centering	
 \includegraphics[width = 1\textwidth]{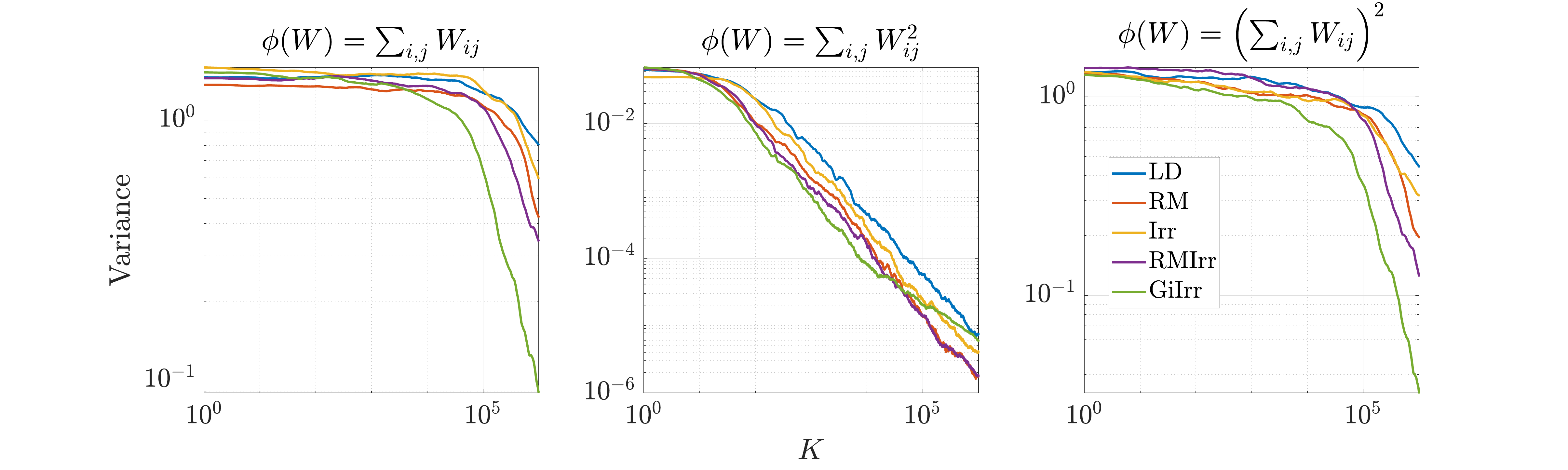}
 	\caption{Variance of running average estimators. For the second moment (middle plot), there is less difference among the samplers as the distribution is quite symmetric, and one can properly estimate the second moment even if the samplers are stuck in a single mode. \texttt{GiIrr} is able to estimate the observable with cross-moments (right plot) better than the other samplers. 
 	}
 	\label{fig:varica}
 \end{figure}

As in the previous example, we simulate the five systems and compute the asymptotic variances of two observables for each system. Each system is simulated independently 100 times up to time $T = 2000$ with $h = 2\times 10^{-5}$. The smaller step size is to account for the additional stiffness irreversible perturbations introduce. Since the true mean of the posterior distribution is unknown, and because standard sampling methods fail to adequately sample from the posterior distribution to get a reasonable estimate for the mean, we only plot the variances \revi{of the selected observables} with respect to $K$ in Figure \ref{fig:varica}. To compute the asymptotic variance, we allot a burn-in time of $T_b = 20$. The observables we estimate are $\phi_1(\W) = \sum_{i,j} \W_{ij}$, $\phi_2(\W) = \sum_{i,j} \W_{ij}^2$, and $\phi_3(\W) = \left(\sum_{i,j}\W_{ij} \right)^2$. The asymptotic variance numbers confirm that the faster mixing observed in the geometry-adapted irreversible perturbation does lead to a better sampling method. The values of the asymptotic variance are reported in Table \ref{table:ica}. The results for the asymptotic variance and variance of $\phi_2$ are somewhat noisy, which is why \texttt{GiIrr} may appear to perform similarly as the other sampling methods.  When estimating the posterior mean and an observable \revi{($\phi_3$)} that includes cross-moments, the geometry-informed irreversible perturbation outperforms standard irreversibility applied to the reversible perturbation.  

\begin{table}[H]
\centering
\begin{tabular}{|l|l|l|l|l|l|l|}
\hline
                      & $\Ex[\text{AVar}_{\phi_1}]$    & $\text{Std}[\text{AVar}_{\phi_1}]$ & $\Ex[\text{AVar}_{\phi_2}]$    & $\text{Std}[\text{AVar}_{\phi_2}]$ & $\Ex[\text{AVar}_{\phi_3}]$    & $\text{Std}[\text{AVar}_{\phi_3}]$ \\ \hline
\texttt{LD}                     & $80.40$ & $23.61$ &$9.445 \times 10^{-4}$ & $3.616\times 10^{-4}$  & $50.17$&$17.52$\\ \hline
\texttt{RM}                     & $53.01$ & $12.22$ & $5.489\times 10^{-4}$ & $1.607\times 10^{-4}$  &$26.75$ &$8.442$    \\ \hline
\texttt{Irr}       & $52.38$ & $17.27$ &$6.854\times 10^{-4}$ & $2.035\times 10^{-4}$     &$27.02$ & $9.134$         \\ \hline
\texttt{RMIrr}  & $39.20$ & $10.36$  &$5.794\times 10^{-4}$ & $1.873\times 10^{-4}$     &$19.47$ & $6.086$           \\ \hline
\texttt{GiIrr}   & $\mathbf{15.50}$ & $4.441$  &$1.253\times 10^{-3}$ & $3.800\times 10^{-4}$     & $\mathbf{6.381}$ &$1.777$       \\ \hline
\end{tabular}
\caption{Asymptotic variance estimates for the ICA example. 
}
\label{table:ica}
\end{table}

\begin{figure}
	\centering
\includegraphics[width = 0.5\textwidth]{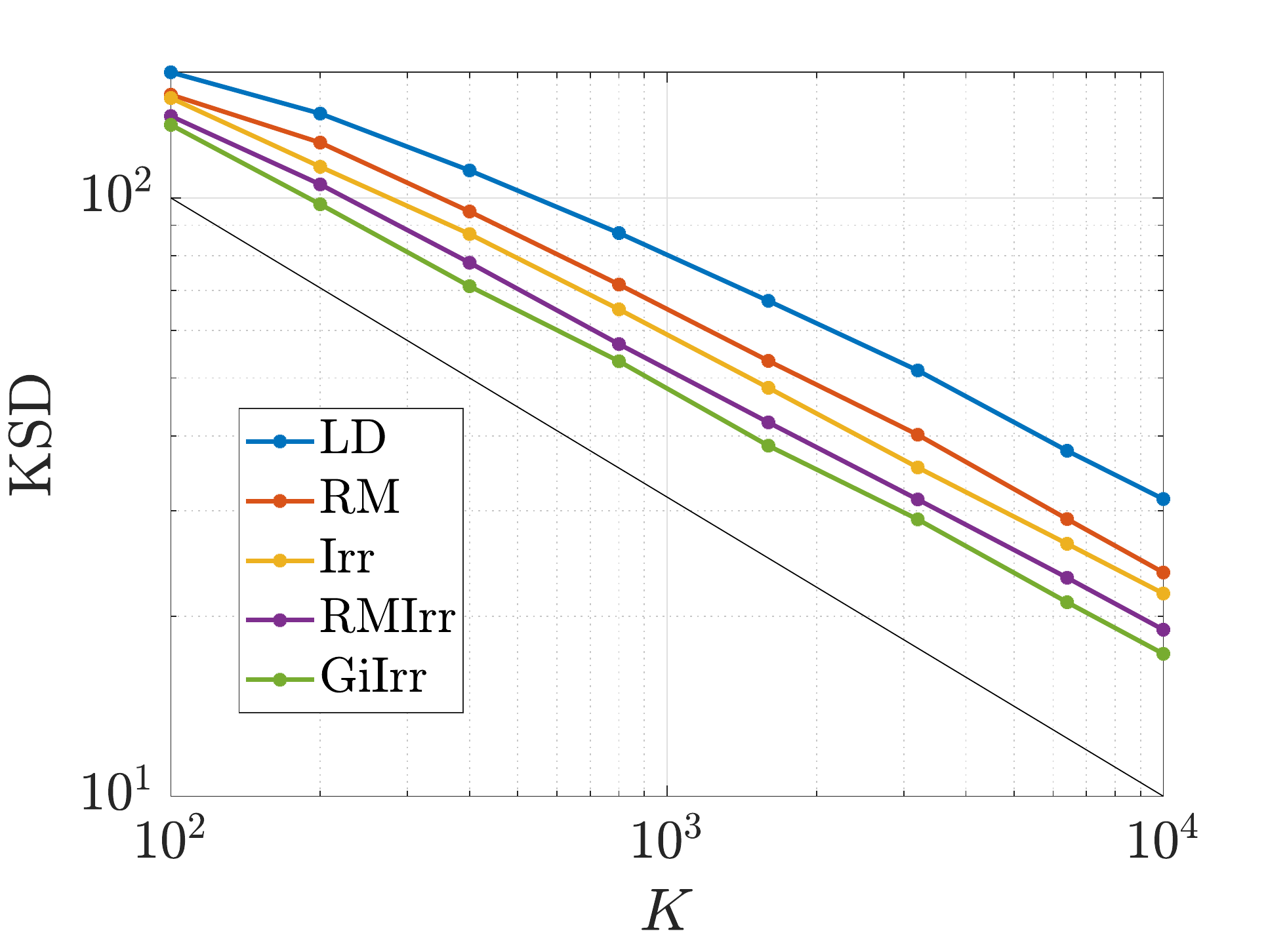}
	\caption{Kernelized Stein discrepancy plot for the ICA example. Black line has slope $-1/2$, which denotes the expected convergence rate. 
	}\label{fig:KSD_ICA}
\end{figure}

\revi{In Figure \ref{fig:KSD_ICA} we plot the convergence of KSD for the ICA example. \texttt{GiIrr} yields lower KSD than the other perturbations in this example, and the theoretical slope of $K^{-1/2}$ is also realized.}

\section{Conclusion}\label{S:Conclusions}
We presented a novel irreversible perturbation, \texttt{GiIrr}, that accelerates the convergence of Langevin dynamics. By introducing an irreversible perturbation that incorporates any given underlying reversible perturbation, which can also be interpreted as defining a Riemannian metric, we have shown through numerical examples that geometry-informed irreversible perturbations outperform those that are not informed as such. In the examples, we found that \texttt{GiIrr} seems to perform best when the target distribution is highly non-Gaussian.

Most of our numerical examples used stochastic gradients to cut down on computational effort in sampling each trajectory. This demonstrates that SGLD can be used in conjunction with irreversibility for practical computations.

We also provided some analysis on how irreversibility interacts with discretization of the SDE systems. Irreversibility introduces additional stiffness into the system, which may lead to additional bias or variance in the estimator. For practical purposes, one can simply choose a small enough step size so that the asymptotic bias and variance are sufficiently small. At the same time, we note an example (see Appendix \ref{sec:appendix}) where the introduction of the irreversible term, once discretized, leads to no improvement in the long term average estimator.

Future work could study the use of novel integrators which circumvent stiffness. For example, \cite{lu2018analysis} uses a multiscale integrator, but it is not readily adapted to the data-driven setting of Bayesian inference. Another direction for future work is to theoretically characterize the performance of the geometry-informed irreversible perturbation and to compare it with that of other perturbations. A starting point for such an analysis could be the general results of \cite{rey2016improving}, in particular the large deviations Theorem 1 together with Propositions 2--4 therein. Preliminary investigation of this direction showed that it is a promising avenue for a theoretical investigation, but non-trivial work and a finer analysis are needed to demonstrate the effects of this class of irreversible perturbations. 
We leave this for future work, as such an analysis is outside the scope of this paper. Our goal in this paper has been to introduce the perturbation and showcase its potential through simulation examples.

\section*{Declarations}
\begin{enumerate}
	\item Funding: BJZ and YMM acknowledge support from the Air Force Office of Scientific Research, Analysis and Synthesis of Rare Events (ANSRE) MURI. KS was partially supported by the National Science Foundation (DMS 1550918, DMS 2107856) and Simons Foundation Award 672441.
	\item Conflicts of interest: Not applicable.
	\item Availability of data and material: Not applicable.
	\item Code availability: Code can be made availble upon request.
\end{enumerate}

\begin{appendices}
\section{The effects of discretization}
\label{sec:appendix}

In this section, we study the effects of discretization in the setting of an irreversibly perturbed Langevin system. Results in full generality are, as yet, elusive; therefore we only consider a Gaussian example, as it still provides insight into how irreversibility interacts with discretization in impacting the asymptotic and finite sample bias and variance of the long term average estimator. While we do not present the results when a stochastic gradient is used, we note that the results are similar and can be easily extended based on what we present here. Recall that,
\begin{align*}
	\A = \frac{1}{2}(\I+\J)(\bGamma_\theta + N \bGamma_X),\,\,\,\, \D = \frac{1}{2}(\I+\J) \left(\bGamma_X \sum_{i = 1}^N X_{i} \right), \text{ where } \J = \delta \begin{bmatrix} 0 & 1 \\ -1 & 0\end{bmatrix}
\end{align*}
For this analysis, all precision matrices are $2\times 2$ scalar matrices. That is, we assume $\bGamma_\theta = \sigma_\theta^{-2}\I$, $\bGamma_X = \sigma_X^{-2}\I$. This is distinct from the example in Section \ref{sec:linear}, since the precision matrices there are diagonal but not scalar. Let $b = \frac{1}{2\sigma_X^2}$ and $S_X = \sum_{k = 1}^N X_i$, so that $\D = b(\I+\J) S_X$.

We summarize our findings here. For fixed discretization size $h$ and scalar precision matrices as defined above, and introducing the irreversible perturbation scaled by $\delta$, we find the following:
\begin{itemize}
	\item The asymptotic bias for linear observables is zero, that is, $\Ex[\theta_\infty] = \mu_p$;
	\item The asymptotic variance for linear observables increases. We found that
	\begin{align}
			\Tr \Var[\theta_\infty] = \frac{2}{2a-ha^2(1+\delta^2)}
	\end{align}
	where $a = 0.5 (1/\sigma_\theta^2 + N/\sigma_X^2)$;
	\item The finite time estimator for the observable $\phi(\theta) = \theta_1 + \theta_2$ has lower bias and variance;
	\item The finite time estimator for the observable $\phi(\theta) = \|\theta\|^2$ has higher bias and variance.
\end{itemize}
We focus on the finite time results and omit the asymptotic results, since the the former case is of more practical interest. The computations related to both are similar.

\paragraph{Finite time analysis: bias for linear observables.}
We study how the magnitude of the irreversibility, characterized by $\delta$, impacts the mean-squared error $\text{MSE} = \Ex\left[ \|\bar{\theta}_K-\mu_p\|^2\right]$ where $\bar{\theta}_K = \frac{1}{K}\sum_{k = 0}^{K-1} \theta_k$. We approach this quantity via its bias-variance decomposition:
\begin{align}
	\text{MSE} = \left\|\Ex[\bar{\theta}_{K}] - \mu _p \right\|^2 + \Tr\text{Var}\left( \bar{\theta}_K\right) .
\end{align}
First, we compute the expected value of the sample average $\Ex\left[\bar{\theta}_K\right] = \frac{1}{K} \sum_{k = 0}^{K-1} \Ex[\theta_k].$ For simplicity, we assume that the initial condition is always $\theta_0 = \mathbf{0}$.
For any $k$, we have
\begin{align*}
	\Ex[\theta_k] &= (\I-h\A)\Ex[\theta_{k-1}] + h\D \\
	& = (\I-h\A)^k \theta_0 + h\sum_{n = 0}^{k-1} (\I-h\A)^n  \D \\
	& =  h(\A h)^{-1}(\I-(\I-h\A)^{k})\D \\
	& =  \A^{-1}\D - \A^{-1}(I-h\A)^k \D.
\end{align*}
This yields
\begin{align*}
	\Ex[\bar{\theta}_K] &= \frac{1}{K}\sum_{k = 0}^{K-1} \left(\A^{-1}\D - \A^{-1}(\I-h\A)^k \D \right) \\
	&= \A^{-1}\D - \frac{1}{K}\A^{-1}(\A h)^{-1}(\I-(\I-h\A)^K)\D.
\end{align*}
Since $\mu_p = \A^{-1}\D$, the bias is
\begin{align*}
	\text{bias} = -\frac{1}{Kh}\A^{-2}(\I-(\I-h\A)^K)\D.
\end{align*}
The norm of the bias can in fact be computed. Note that $\A^2 = (1+\delta^2)a^2\I$ and we have
\begin{align*}
	\|\text{bias}\|^2 &= \frac{1}{K^2h^2} \D^T (\I-(\I-h\A^T)^K)\A^{-2T}\A^{-2}(\I-(\I-h\A)^K) \D \\
	& = \frac{1}{K^2h^2a^4(1+\delta^2)^2}\D^T(\I-(\I-h\A^T)^K)(\I-(\I-h\A)^K) \D \\
	& = \frac{b^2}{K^2h^2a^4(1+\delta^2)^2}S_X^T(\I+\J)^T(\I-(\I-\A^Th)^K)(\I-(\I-h\A)^K) (\I+\J) S_X.
\end{align*}
The inner matrix can be computed. Since each matrix above is simultaneously diagonalizable, we only need to consider the eigenvalues of each of the above matrices. Note that $\I+\J$ is a normal matrix, so we may write the eigenvalue decomposition $\I+\J = \mathbf{PDP}^*$, where $^*$ denotes conjugate transpose, $\mathbf{Q} = \text{diag}(1+i\delta,1-i\delta)$, and
\begin{align*}
	\mathbf{P} = \frac{1}{\sqrt{2}}\begin{bmatrix}
		1 & 1 \\ i & -i
	\end{bmatrix}
\end{align*}
is orthogonal. Now note that
\begin{align*}
\I-(\I-h\A)^K = \mathbf{P} \begin{bmatrix}
		1-(1-ah(1+i\delta))^K & 0 \\0 & 1-(1-ah(1-i\delta))^K
	\end{bmatrix} \mathbf{P}^*,
\end{align*}
which implies
\begin{align*}
	(\I-(\I-h\A^T)^K)(\I-(\I-h\A)^K)= |1-(1-ah(1+i\delta)^K)|^2\I.
\end{align*}
Using the fact that $(\I+\J)^T(\I+\J) = (1+\delta^2)I$, and we have the following
\begin{align}
	\|\text{bias}\|^2 &= \frac{b^2}{K^2h^2a^4(1+\delta^2)} |1-(1-a(1+i\delta)h)^K|^2 \|S_X\|.
\end{align}
To simplify further, we write $1-a(1+i\delta)h = re^{i\theta}$ where $r^2 = (1-ah)^2 + \delta^2a^2h^2$, and $\tan \theta = \delta a h/ (1-ah)$. Then we obtain
\begin{align*}
	\|\text{bias}\|^2 &= \frac{b^2}{K^2h^2a^4(1+\delta^2)} |1-r^Ke^{i\theta K}|^2 \| S_X \|^2 \\
	& = \frac{b^2}{K^2h^2a^4(1+\delta^2)} (1+r^{2K}-2r^K \cos K\theta )\| S_X \|^2.
\end{align*}
We know that $r<1$, since otherwise, the numerical scheme would be unstable. It is easy to see that for large, but not infinite, $K$, the bias decays as $\mathcal{O}(1/(Kh\sqrt{1+\delta^2}))$, so the introduction of irreversibility decreases the constant in front of the expression and therefore slightly improves the convergence of the bias.

\paragraph{Finite time analysis: variance for linear observables.}
For simplicity, we assume $\theta_0 = 0$. We compute $\text{TrVar}(\bar{\theta}_K)$. We begin with
\begin{align*}
	\Tr\text{Var}(\bar{\theta}_K) = \Tr\Ex[\bar{\theta}_K\bar{\theta}_K^T] - \Tr\Ex[\bar{\theta}_K]\Ex[\bar{\theta}_K]^T
\end{align*}
and compute these terms separately. It is difficult to surmise a relationship between $\delta$ and $\Tr\Var(\bar{\theta}_K)$ even with exact formulas, so we appeal to plots of the expressions to see that the variance decreases with irreversibility. We computed $\Ex[\bar{\theta}_K]$ in the previous section.

With the observation that
\begin{align*}
	\A^{-2}(\I-(\I-h\A)^K) \Ex[\D] = \frac{b}{a^2}\mathbf{PQ'P}^*\Ex[S_X]
\end{align*}
where
\begin{align*}
	\mathbf{Q}' = \begin{bmatrix}
		\frac{1-(1-ah(1+i\delta))^K}{1+i\delta} & 0 \\ 0 & \frac{1-(1-ah(1-i\delta))^K}{1-i\delta},
	\end{bmatrix}
\end{align*}
and $\mathbf{P}$ is defined in the previous section. We compute that
\begin{align}
	\Tr\Ex[\bar{\theta}_K] \Ex[\bar{\theta}_K]^T = \|\mu_p\|^2 + \|\text{bias}\|^2 - \frac{2b^2}{Kha^3(1+\delta^2)}\text{Re}\{(1-i\delta)(1-ah(1+i\delta))^K \} \|\Ex[S_X]\|^2.
\end{align}

The other term is more complicated and needs to be approached more gingerly. Observe that
\begin{align*}
	\Tr\Ex[\bar{\theta}_K\bar{\theta}_K^T ] = \frac{1}{K^2} \sum_{i,j = 1}^K \Tr\Ex[\theta_i\theta_j^T] = \frac{1}{K^2} \left(\sum_{i = 0}^{K-1} \Tr\Ex[\theta_i\theta_i^T] + 2\sum_{i<j = 0}^{K-1} \Tr\Ex[\theta_i\theta_j^T] \right).
\end{align*}
We take each term individually.
To compute $\Ex[\theta_k\theta_k^T]$, it is actually better to consider the covariance matrix of $\theta_k$, $\Sigma_{k} = \Ex[\theta_k\theta_k^T] - \Ex[\theta_k]\Ex[\theta_k]^T$.

We first compute
 \begin{align*}
 	\Ex[ \theta_{k} \theta_{k}^T ] = &(\I-h\A) \Ex[\theta_{k-1}\theta_{k-1}^T] (\I-h\A)^T + h^2 \D \D^T + h \I \\&+ (\I-h\A) \Ex[\theta_{k-1}] \D^T h + h\D \Ex[ \theta_{k-1}^T](\I-h\A)^T \\
 	\Ex[\theta_{k}]\Ex[\theta_{k}]^T =& (\I-h\A) \Ex[\theta_{k-1}]\Ex[\theta_{k-1}]^T (\I-h\A)^T + \D\Ex[\theta_{k-1}]^T (\I-h\A)^T \\& + (\I-h\A) \Ex[\theta_{k-1}]\D+ h^2\D\D^T
 \end{align*}
 which imply the following recurrence relation. Assuming $\Sigma_0 = \mathbf{0}$, we have
 \begin{align*}
 	\Sigma_{k} = (\I-h\A) \Sigma_{k-1}(\I-h\A)^T + h\I &= h \sum_{n = 0}^{k-1}\left( (\I-h\A)(\I-h\A)^{T}\right)^n \\
	&= h\sum_{n = 0}^{k-1}(\I-(\A+\A^T)h + h^2\A\A^T)^n  \\ &= ((\A+\A^T)-h^2\A\A^T)^{-1}(\I-(\I-(\A+\A^T)h+\A\A^T h^2)^k).
\end{align*}
Let $s = 1-2ah + h^2a^2(1+\delta^2)$, then, by recalling that $\A+\A^T = 2a\I$ and $ \A\A^T = a^2(1+\delta^2)\I$, the above sum is equal to	$\frac{1-s^k}{1-s} h\I$.
Therefore,
\begin{align}
	\Tr \Sigma_k = \frac{2h(1-s^k)}{1-s}.
\end{align}
Meanwhile note that
\begin{align*}
	\Ex[\theta_k] &= \mu_p -\A^{-1} (\I-h\A)^k \D.
\end{align*}
Therefore,
\begin{align*}
	\Tr\Ex[\theta_k]\Ex[\theta_k]^T = \Ex[\theta_k]^T\Ex[\theta_k] &= \|\mu_p\|^2 + \D^T(\I-h\A^T)^k \A^{-T}\A^{-1}(\I-h\A)^k\D-2\mu_p^T\A^{-1}(\I-h\A)^k \D \\
	& = \|\mu_p\|^2 + \frac{s^k b^2}{a^2(1+\delta^2)}\|S_X\|^2-2\mu_p^T\A^{-1}(\I-h\A)^k \D.
\end{align*}
We now take the sum for each expression from $k = 0$ to $K-1$. We have
\begin{align*}
	\sum_{i =0}^{K-1} \Tr\Sigma_i &= 2h\left(\frac{K}{1-s}- \sum_{i = 0}^{K-1}\frac{s^i}{1-s}\right)  \\
	& = 2h\left(\frac{K}{1-s}- \frac{1-s^K}{(1-s)^2}\right) 
\end{align*}
and
\begin{align*}
	\sum_{i = 0}^{K-1}\Ex[\theta_i]^T\Ex[\theta_i]
	& = K\|\mu_p\|^2 + \frac{(1-s^K) b^2}{a^2(1+\delta^2)(1-s)}\|S_X\|^2-2\mu_p^T\A^{-1}(h\A)^{-1}(\I-(\I-h\A)^K) \D\\
	& = K\|\mu_p\|^2 + \frac{(1-s^K) b^2}{a^2(1+\delta^2)(1-s)}\|S_X\|^2-2\mu_p^Th^{-1}\A^{-2}(\I-(\I-h\A)^K) \D.
\end{align*}

For the cross-terms, observe that we may write
\begin{align}
	\sum_{i<j = 0}^{K-1} \theta_i\theta_j^T = \sum_{i = 0}^{K-1} \theta_i \sum_{j = i+1}^{K-1} \theta_j^T
\end{align}
which can be simplified further. First note that
\begin{align*}
	\theta_j &= (\I-h\A) \theta_{j-1} + \D h + \sqrt{h}\xi_{j-1} \\
	& = (\I-h\A)^{j-i} \theta_i + h\sum_{n = 0}^{j-1-i} (\I-h\A)^{n} \D + \sqrt{h} \sum_{n = 0}^{j-i-1} (\I-h\A)^{n} \xi_{j-1-n}.
\end{align*}
Plugging this expression into the double sum above, we have
\begin{align}
	\sum_{i =0}^{K-1} \theta_i \sum_{j = i+1}^{K-1}\left[\theta_i^T(\I-\A^Th)^{j-i} + h \sum_{n = 0}^{j-1-i} \D^T(\I-\A^Th)^n + \sqrt{h}\sum_{n = 0}^{j-i-1}\xi_{j-1-n}^T(\I-\A^Th)^n \right].
\end{align}
Taking expectations, we have
\begin{align*}
	& \sum_{i = 0}^{K-1}  \sum_{j = i+1}^{K-1} \left[\Ex[\theta_i\theta_i^T](\I-h\A^T)^{j-i} + h\Ex[\theta_i]\D^T \sum_{n = 0}^{j-1-i} (\I-h\A^T)^n  \right] \\
	= & \sum_{i = 0}^{K-1}  \sum_{j = i+1}^{K-1} \left[\Ex[\theta_i\theta_i^T](\I-h\A^T)^{j-i} + h\Ex[\theta_i]\D^T (\A^Th)^{-1}(\I-(\I-\A^Th)^{j-i})  \right].
\end{align*}
Carrying out the computation for the first term, we have
\begin{align*}
	F = \sum_{i = 0}^{K-1} \Ex[\theta_i\theta_i^T] \sum_{j = i+1}^{K-1} (\I-\A^Th)^{j-i} = \sum_{i = 0}^{K-1} \Ex[\theta_i\theta_i^T] (\I-\A^Th)(\A^Th)^{-1}(\I-(\I-\A^Th)^{K-1-i}).
\end{align*}
For the second term we have,
\begin{align*}
	&\sum_{i = 0}^{K-1} \Ex[\theta_i]\mu_p^T \sum_{j = i+1}^{K-1} (\I-(\I-\A^Th)^{j-i})  \nonumber\\
&\qquad= \sum_{i = 0}^{K-1} \Ex[\theta_i] \mu_p^T \left[(K-1-i)\I - (\I-\A^Th)(\A^Th)^{-1} (\I-(\I-\A^Th)^{K-1-i})  \right].
\end{align*}
The summations are difficult to compute precisely, so we compute them by direct evaluation instead.  For simplicity, we assume that $\mu_p = \Ex[S_X] = [0,0]^T$, $\sigma_X$ and $\sigma_\theta$ are chosen such that $a=1$. For this scenario, the bias is zero and only the variance contributes to the MSE. The variance is
\begin{align*}
	\Tr[\Var\overline{\theta}_K] = \frac{1}{K^2}\left( 2h\left(\frac{K}{1-s} - \frac{1-s^K}{(1-s)^2} \right) + 2\Tr F\right)
\end{align*}
where $\Ex[\theta_i\theta_i^T] = \Sigma_i = \frac{1-s^i}{1-s}h\I.$

\begin{figure}
\centering
\includegraphics[width = 0.4\textwidth]{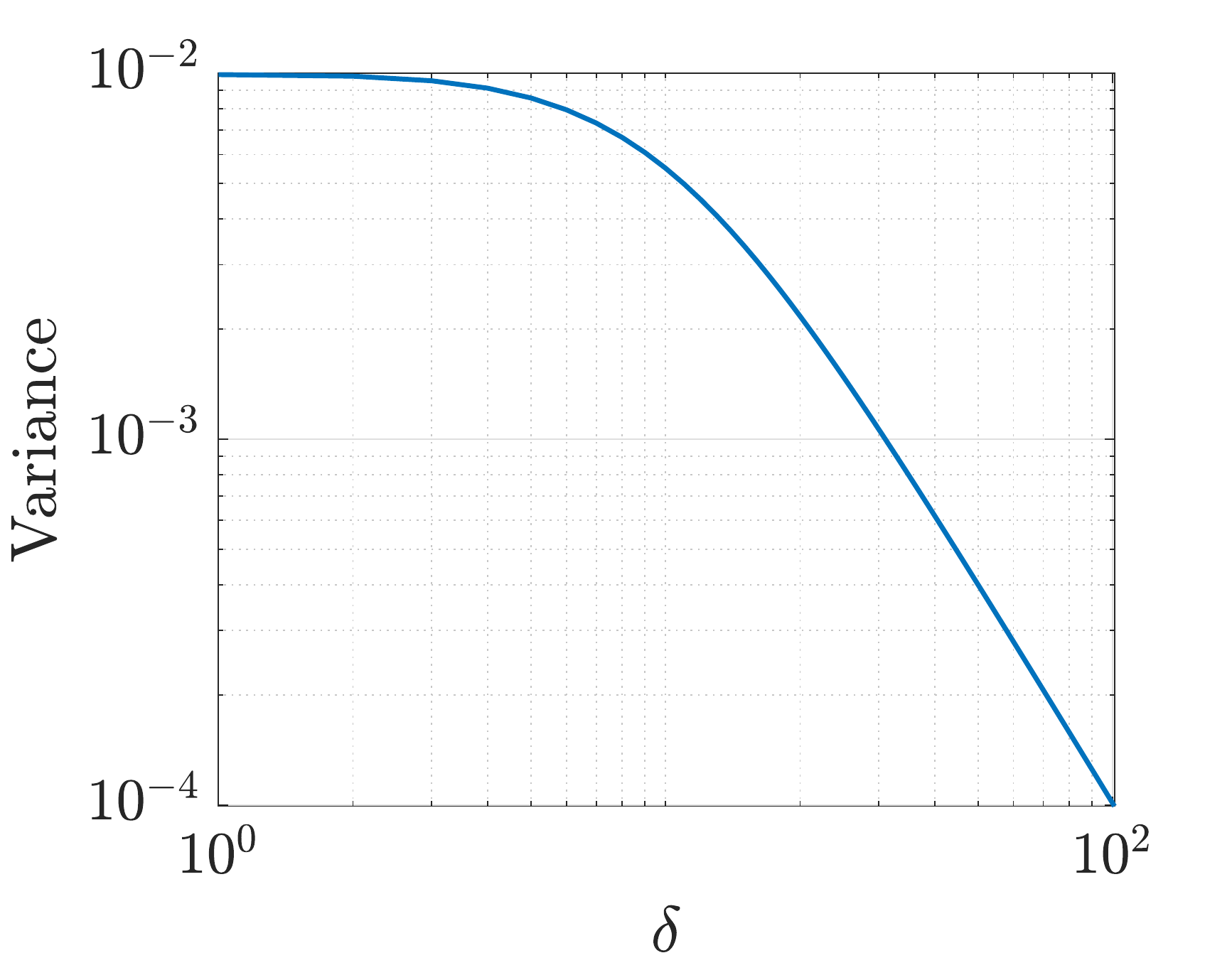}
	\caption{Variance for different $\delta$, fixed $h$.}
	\label{fig:biasvardirecteval}
\end{figure}

In Figure \ref{fig:biasvardirecteval} we plot the variance for varying choices of $\delta$. In this plots, $h = 0.001$, $K = 2\times 10^5$, and $\delta$ varies between zero and ten. We can clearly see that strengthening the irreversible perturbation leads to improvement of the squared bias and variance of the long term average estimator.

\paragraph{Finite sample analysis for the quadratic  observable $\phi(\theta) = \|\theta\|^2$.}

\begin{figure}
\centering
\includegraphics[width = \textwidth]{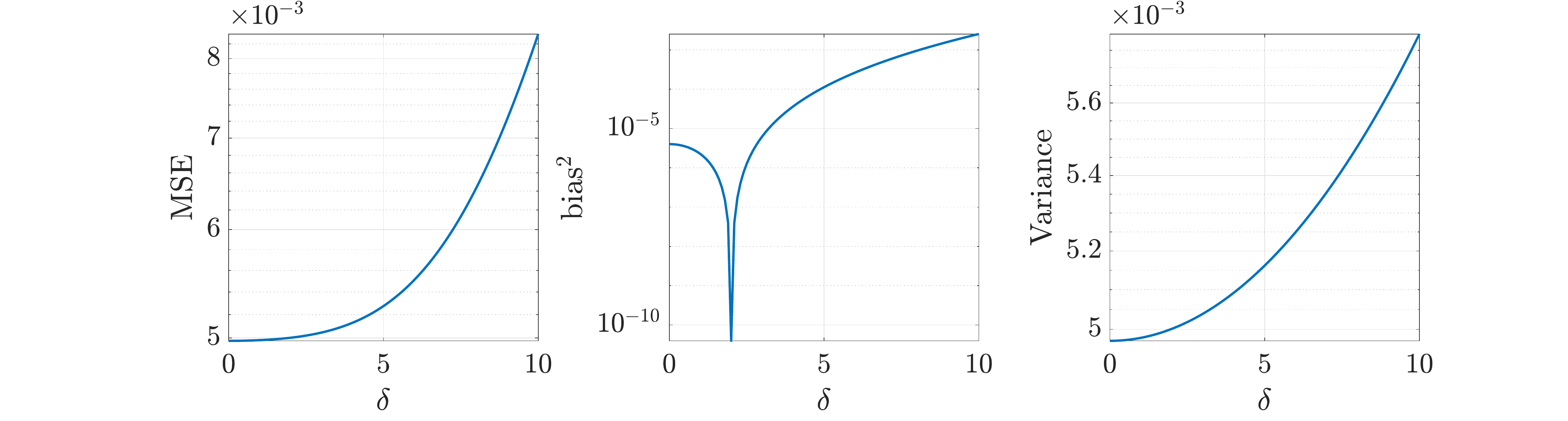}
	\caption{(Squared) bias and variance of $\phi(\theta) = \|\theta\|^2$ for varying levels of irreversibility.}\label{fig:secobs}
\end{figure}

The previous finite sample results for the observable $\phi(\theta) = \theta_1 + \theta_2$ suggests that both the bias and variance of the long term average estimator goes down with a larger irreversible term. In this section, we show that this is actually a special case, and that when the observable is not linear, then the bias and variance may increase. We analyze the bias and variance of the long term average estimator of the observable $\phi(\theta) = \|\theta\|^2$. Define
\begin{align}
	\overline{\phi} = \int \phi(\theta) \pi(\theta) \de\theta, \;\;\;\; \overline{\phi}_K = \frac{1}{K}\sum_{k = 0}^{K-1} \phi(\theta_k).
\end{align}
  As before, we assume that $\mu_p = [0,0]^T$, $\Ex[S_X] = \mathbf{0}$, and $\sigma_x$ and $\sigma_\theta$ are chosen such that $a=1$. We compute $|\Ex\overline{\phi}_K - \overline{\phi}|^2$ and $\Var \overline{\phi}_k$ and see how they vary with $\delta$. From previous computations, we can show that
	\begin{align}
	\Ex\overline{\phi}_K = 2h \left(\frac{1}{1-s} - \frac{1-s^K}{K(1-s)^2} \right),
	\end{align}
	where $s = 1-2ah + a^2h^2(1+\delta^2)$. Given this, the only term left to compute is the variance of the second moment of this observable:
\begin{align}
	\Ex\left[\left( \overline{\phi}_K\right)^2 \right] &= \frac{1}{K^2} \Ex\left[\left(\sum_{k = 0}^{K-1} \theta_k^T\theta_k \right)^2 \right] \nonumber\\
	& = \frac{1}{K^2}\sum_{k = 0}^{K-1} \Ex\left[(\theta_k^T\theta_k)^2 \right] + \frac{2}{K^2}\Ex \sum_{k = 0}^{K-1} \sum_{l = k+1}^{K-1} (\theta_k^T\theta_k) (\theta_l^T \theta_l).
\end{align}
To compute the first sum, consider the following:
\begin{align*}
	\theta_k^T\theta_k &= \theta_{k-1}^T(\I-h\A)^T(\I-h\A)\theta_{k-1} + h\xi_{k-1}^T\xi_{k-1} + 2\sqrt{h} \theta_{k-1}^T(\I-h\A)^T \xi_{k-1}
\end{align*}
and so we have
\begin{align*}
	(\theta_k^T \theta_k)^2 =& s^2(\theta_{k-1}^T\theta_{k-1})^2 + h^2 (\xi_{k-1}^T \xi_{k-1})^2 + 4h (\xi_{k-1}^T(\I-h\A)\theta_{k-1})^2+2s\theta_{k-1}^T\theta_{k-1} h \xi_{k-1}^T \xi_{k-1} \\
	&+ 4 s\sqrt{h}(\theta_{k-1}^T\theta_{k-1}) \theta_{k-1}^T(\I-h\A)^T\xi_{k-1}+4h^{3/2}(\xi_{k-1}^T\xi_{k-1})\theta_{k-1}^T(\I-h\A)^T\xi_{k-1}.
\end{align*}
Taking the expectation, we have
\begin{align*}
	\Ex[(\theta_k^T\theta_k)^2] = &s^2 \Ex[(\theta_{k-1}^T\theta_{k-1})^2] + h^2 \Ex[(\xi_{k-1}^T\xi_{k-1})^2] + 4h\Ex\left[(\xi_{k-1}^T(\I-h\A)\theta_{k-1})^2 \right] \\&+2sh\Ex\left[(\theta_{k-1}^T\theta_{k-1})(\xi_{k-1}^T\xi_{k-1}) \right].
\end{align*}
After simplifying, we arrive at the following recurrence relation:
\begin{align}
	\Ex\left[(\theta_k^T\theta_k)^2 \right] = s^2 \Ex\left[ (\theta_{k-1}^T\theta_{k-1})^2\right] + 8h^2 + 8sh \Ex\left[\theta_{k-1}^T\theta_{k-1}\right].
\end{align}
Let $\beta_k = \Ex[(\theta_k^T\theta_k)^2]$, $\zeta_k = 8sh\Ex[\theta_k^T\theta_k]$, and $\kappa = 8h^2$. We have the following recurrence, which we solve
\begin{align*}
	\beta_k &= s^2 \beta_{k-1} + \zeta_{k-1} + \kappa \\
	& = s^{2k} \beta_0 + \sum_{n = 0}^{k-1} s^{2n} \zeta_{k-n-1} + \sum_{n = 0}^{k-1} s^{2n} \kappa.
\end{align*}
From previous for the term $\zeta_k$, we have
\begin{align*}
	\beta_k &= \sum_{n = 0}^{k-1} s^{2n} 8sh \cdot 2h \frac{1-s^{k-n-1}}{1-s} + \kappa \frac{1-s^{2k}}{1-s^2} \\
	&= \frac{16sh^2}{1-s} \sum_{n = 0}^{k-1} (s^{2n} - s^{k+n-1}) + \frac{8h^2(1-s^{2k})}{1-s^2} \\
	& = \frac{16sh^2}{1-s} \left(\frac{1-s^{2k}}{1-s^2} - s^{k-1} \frac{1-s^k}{1-s} \right) + \frac{8h^2(1-s^{2k})}{1-s^2}.
\end{align*}
Next we compute the summation of the cross terms. Define $R_k$ such that
\begin{align}
	\sum_{k = 0}^{K-1} R_k = \sum_{k = 0}^{K-1} \sum_{l = k+1}^{K-1} \Ex\left[(\theta_k^T\theta_k)(\theta_l^T\theta_l)\right].
\end{align}
We write
\begin{align*}
	\theta_l^T \theta_l& = s\theta_{l-1}^T\theta_{l-1} + h \xi_{l-1}^T\xi_{l-1} + 2\sqrt{h} \xi_{l-1}^T (\I-h\A) \theta_{l-1} \\
	& = s^{l-k} \theta_k^T\theta_k + \sum_{n = 0}^{l-k-1} hs^n \xi_{l-n-1}^T\xi_{l-n-1} + 2\sqrt{h} s^n \xi_{l-n-1}^T(\I-h\A) \theta_{l-n-1}.
\end{align*}
This implies that
\begin{align*}
	R_k &= \sum_{l = k+1}^{K-1} \left(s^{l-k}\Ex[ (\theta_k^T\theta_k)^2]	+ \sum_{n = 0}^{l-k-1} 2h s^n \Ex[\theta_k^T\theta_k] \right)\\
	& = \sum_{l = k+1}^{K-1} \beta_ks^{l-k} + 2h\Ex[\theta_k^T\theta_k]\sum_{n = 0}^{l-k-1} s^n  \\
	& = \sum_{l = k+1}^{K-1} \beta_k s^{l-k} + 2h\Ex[\theta_k^T\theta_k] \frac{1-s^{l-k}}{1-s} \\
	& = \beta_k \sum_{l = k+1}^{K-1} s^{l-k} + \frac{2h\Ex[\theta_k^T\theta_k]}{1-s} \sum_{l = k+1}^{K-1} 1-s^{l-k} \\
	& = \beta_k \frac{s-s^{K-k}}{1-s} + \frac{2h\Ex[\theta_k^T\theta_k]}{1-s}\left(K-1-k - \frac{s-s^{K-k}}{1-s} \right).
\end{align*}
To summarize, we have
\begin{align}
	\Ex\left[\left(\overline{\phi}_K \right)^2 \right] = \frac{1}{K^2} \sum_{k = 0}^{K-1}\left( \beta_k +2 R_k \right),
\end{align}
the squared bias is $(\Ex\overline{\phi}_K - 1)^2$ and the variance is $\Ex[(\overline{\phi}_K)^2] - \Ex[\overline{\phi}_K]^2$. These expressions are not simplifiable easily, so we plot these expressions and study their trends. In Figure \ref{fig:secobs}, we plot the squared bias and variance for fixed $h$ and $K$ and varying $\delta$. In these plots, $h = 0.001$, $K = 2\times 10^5$, and $\delta$ varies between zero and ten. Notice that for these choices, both the squared bias and variance increases as $\delta$ grows, showing that for irreversibility provides no benefit, and in fact, harms the performance of the standard estimator.

\begin{remark}\label{R:IrrevNotImprovement}
When discretization is considered, sampling properties in this simple example in which the drift is proportional to the state cannot be improved upon using irreversibility. Let us now explain this phenomenon from a theoretical point of view. In \cite{rey2016improving}, the authors show that adding an irreversible perturbation to the generator of the diffusion process may decrease the spectral gap, and will never increase it. They further prove that in the continuous case, decreasing the spectral gap then decreases the asymptotic variance. However this improvement is not strict, that is, irreversibility is only guaranteed to not increase the spectral gap.

Meanwhile, in \cite{lelievre2013optimal}, the authors consider irreversibility only in the context of linear systems, and rigorously study optimal irreversible perturbations that accelerate convergence to the invariant distribution. Their results show that when the drift matrix is proportional to the identity matrix, the spectral gap cannot be widened. Proposition 4 in \cite{lelievre2013optimal} shows that the leading nonzero eigenvalue of the irreversibly perturbed drift matrix is bounded above by the leading nonzero eigenvalue of the original drift matrix and below by the trace of the original drift matrix over the dimension of the state space. The lower bound is then the optimal spectral gap. For a drift matrix that is a multiple of the identity, the upper and lower bounds are the same, which implies that the spectral gap can never decrease from its original value in the continuous case. After factoring in discretization, the irreversible perturbation increases stiffness of the system, which contributes to increased bias and variance in the resulting estimator.
\end{remark}

\end{appendices}

\bibliographystyle{plain}
\bibliography{bibliotheque}

\begin{thebibliography}{10}

\bibitem{amari1996new}
Shun-ichi Amari, Andrzej Cichocki, and Howard~Hua Yang.
\newblock A new learning algorithm for blind signal separation.
\newblock In {\em Advances in {N}eural {I}nformation {P}rocessing {S}ystems},
  pages 757--763. Morgan Kaufmann Publishers, 1996.

\bibitem{asmussen2007stochastic}
S{\o}ren Asmussen and Peter~W Glynn.
\newblock {\em Stochastic simulation: algorithms and analysis}, volume~57.
\newblock Springer Science \& Business Media, 2007.

\bibitem{Bierkens}
Joris Bierkens.
\newblock Non-reversible {M}etropolis-{H}astings.
\newblock {\em Statistics and Computing}, 26:1213--1228, 2016.

\bibitem{BrosseMoulinesDurmus2018}
Nicolas Brosse, Alain Durmus, and {\'E}ric Moulines.
\newblock The promises and pitfalls of stochastic gradient {L}angevin dynamics.
\newblock In {\em NeurIPS 2018 (Advances in Neural Information Processing
  Systems 2018)}, 2018.

\bibitem{DiaconisHolmesNeal2010}
P~Diaconis, S~Holmes, and R~Neal.
\newblock Analysis of a nonreversible {M}arkov chain sampler.
\newblock {\em Annals of Applied Probability}, 10:726--752, 2010.

\bibitem{DuncanPavliotisZygalakis2017}
A.B. Duncun, G.A. Pavliotis, and K.C. Zygalakis.
\newblock Nonreversible {L}angevin samplers: Splitting schemes, analysis and
  implementation.
\newblock {\em arXiv preprint:1701.04247}, 2017.

\bibitem{durmus2019high}
Alain Durmus and Eric Moulines.
\newblock High-dimensional {B}ayesian inference via the unadjusted {L}angevin
  algorithm.
\newblock {\em Bernoulli}, 25(4A):2854--2882, 2019.

\bibitem{franke2010behavior}
Brice Franke, C-R Hwang, H-M Pai, and S-J Sheu.
\newblock The behavior of the spectral gap under growing drift.
\newblock {\em Transactions of the American Mathematical Society},
  362(3):1325--1350, 2010.

\bibitem{Ganguly}
Arnab Ganguly and P~Sundar.
\newblock Inhomogeneous functionals and approximations of invariant
  distribution of ergodic diffusions: Error analysis through central limit
  theorem and moderate deviation asymptotics.
\newblock {\em Stochastic Processes and their Applications}, 133(C):74--110,
  2021.

\bibitem{gershman2012nonparametric}
Samuel~J. Gershman, Matthew~D. Hoffman, and David~M. Blei.
\newblock Nonparametric variational inference.
\newblock In {\em Proceedings of the 29th International Conference on
  International Conference on Machine Learning}, ICML'12, pages 235\--242,
  Madison, WI, USA, 2012. Omnipress.

\bibitem{girolami2011riemann}
Mark Girolami and Ben Calderhead.
\newblock Riemann manifold {L}angevin and {H}amiltonian {M}onte {C}arlo
  methods.
\newblock {\em Journal of the Royal Statistical Society: Series B (Statistical
  Methodology)}, 73(2):123--214, 2011.

\bibitem{gorham2019measuring}
Jackson Gorham, Andrew~B Duncan, Sebastian~J Vollmer, and Lester Mackey.
\newblock Measuring sample quality with diffusions.
\newblock {\em The Annals of Applied Probability}, 29(5):2884--2928, 2019.

\bibitem{gorham2015measuring}
Jackson Gorham and Lester Mackey.
\newblock Measuring sample quality with stein's method.
\newblock {\em Advances in Neural Information Processing Systems}, 28, 2015.

\bibitem{gorham2017measuring}
Jackson Gorham and Lester Mackey.
\newblock Measuring sample quality with kernels.
\newblock In {\em International Conference on Machine Learning}, pages
  1292--1301. PMLR, 2017.

\bibitem{hu2020non}
Yuanhan Hu, Xiaoyu Wang, Xuefeng Gao, Mert G{\"u}rb{\"u}zbalaban, and Lingjiong
  Zhu.
\newblock Non-convex optimization via non-reversible stochastic gradient
  {L}angevin dynamics.
\newblock {\em arXiv preprint arXiv:2004.02823}, 2020.

\bibitem{hwang1993accelerating}
Chii-Ruey Hwang, Shu-Yin Hwang-Ma, and Shuenn-Jyi Sheu.
\newblock Accelerating {G}aussian diffusions.
\newblock {\em The Annals of Applied Probability}, pages 897--913, 1993.

\bibitem{hwang2005accelerating}
Chii-Ruey Hwang, Shu-Yin Hwang-Ma, and Shuenn-Jyi Sheu.
\newblock Accelerating diffusions.
\newblock {\em Annals of Applied Probability}, 15(2):1433--1444, 2005.

\bibitem{izzatullah2020bayesian}
Muhammad Izzatullah, Ricardo Baptista, Lester Mackey, Youssef Marzouk, and
  Daniel Peter.
\newblock Bayesian seismic inversion: Measuring langevin mcmc sample quality
  with kernels.
\newblock In {\em SEG International Exposition and Annual Meeting}. OnePetro,
  2020.

\bibitem{lelievre2013optimal}
Tony Lelievre, Francis Nier, and Grigorios~A Pavliotis.
\newblock Optimal non-reversible linear drift for the convergence to
  equilibrium of a diffusion.
\newblock {\em Journal of Statistical Physics}, 152(2):237--274, 2013.

\bibitem{LiuLeeJordan2016}
Q.~Liu, J.~Lee, and M.~Jordan.
\newblock A kernelized {Stein} discreprancy for goodness-of-fit tests.
\newblock {\em In Proc. of 33rd ICML}, 48:276--284, 2016.

\bibitem{livingstone2014information}
Samuel Livingstone and Mark Girolami.
\newblock Information-geometric {M}arkov chain {M}onte {C}arlo methods using
  diffusions.
\newblock {\em Entropy}, 16(6):3074--3102, 2014.

\bibitem{lu2018analysis}
Jianfeng Lu and Konstantinos Spiliopoulos.
\newblock Analysis of multiscale integrators for multiple attractors and
  irreversible {L}angevin samplers.
\newblock {\em Multiscale Modeling \& Simulation}, 16(4):1859--1883, 2018.

\bibitem{Ma_recipe2015}
Yi-An Ma, Tianqi Chen, and Emily~B. Fox.
\newblock A complete recipe for stochastic gradient {MCMC}.
\newblock {\em NIPS'15: Proceedings of the 28th International Conference on
  Neural Information Processing Systems}, 2:2917--2925, 2015.

\bibitem{ottobre2019optimal}
Michela Ottobre, Natesh~S Pillai, and Konstantinos Spiliopoulos.
\newblock Optimal scaling of the {MALA} algorithm with irreversible proposals
  for {G}aussian targets.
\newblock {\em Stochastics and Partial Differential Equations: Analysis and
  Computations}, pages 1--51, 2019.

\bibitem{pavliotis2014stochastic}
Grigorios~A Pavliotis.
\newblock {\em Stochastic processes and applications: diffusion processes, the
  Fokker-Planck and {L}angevin equations}, volume~60.
\newblock Springer, 2014.

\bibitem{rey2015irreversible}
Luc Rey-Bellet and Konstantinos Spiliopoulos.
\newblock Irreversible {L}angevin samplers and variance reduction: a large
  deviations approach.
\newblock {\em Nonlinearity}, 28(7):2081, 2015.

\bibitem{rey2015variance}
Luc Rey-Bellet and Konstantinos Spiliopoulos.
\newblock Variance reduction for irreversible {L}angevin samplers and diffusion
  on graphs.
\newblock {\em Electronic Communications in Probability}, 20, 2015.

\bibitem{rey2016improving}
Luc Rey-Bellet and Konstantinos Spiliopoulos.
\newblock Improving the convergence of reversible samplers.
\newblock {\em Journal of Statistical Physics}, 164(3):472--494, 2016.

\bibitem{roberts1996exponential}
Gareth~O Roberts and Richard~L. Tweedie.
\newblock Exponential convergence of {L}angevin distributions and their
  discrete approximations.
\newblock {\em Bernoulli}, 2(4):341--363, 1996.

\bibitem{teh2016consistency}
Yee~Whye Teh, Alexandre~H Thiery, and Sebastian~J Vollmer.
\newblock Consistency and fluctuations for stochastic gradient {L}angevin
  dynamics.
\newblock {\em Journal of Machine Learning Research}, 17, 2016.

\bibitem{vollmer2016exploration}
Sebastian~J Vollmer, Konstantinos~C Zygalakis, and Yee~Whye Teh.
\newblock Exploration of the (non-) asymptotic bias and variance of stochastic
  gradient {L}angevin dynamics.
\newblock {\em The Journal of Machine Learning Research}, 17(1):5504--5548,
  2016.

\bibitem{welling2011bayesian}
Max Welling and Yee~W Teh.
\newblock Bayesian learning via stochastic gradient {L}angevin dynamics.
\newblock In {\em Proceedings of the 28th {I}nternational {C}onference on
  {M}achine {L}earning (ICML-11)}, pages 681--688. Citeseer, 2011.

\bibitem{xifara2014langevin}
Tatiana Xifara, Chris Sherlock, Samuel Livingstone, Simon Byrne, and Mark
  Girolami.
\newblock {L}angevin diffusions and the {M}etropolis-adjusted {L}angevin
  algorithm.
\newblock {\em Statistics \& Probability Letters}, 91:14--19, 2014.

\end{thebibliography}

\end{document}